\documentclass[final,12pt]{article}

\usepackage[usenames,dvipdfmx]{color}
\usepackage{setspace,amsfonts,comment,amsmath,amsthm,amssymb,amsxtra,graphicx}
\usepackage[pdftex,pagebackref]{hyperref}
\usepackage{mathpazo}

\usepackage{threeparttable}
\usepackage{booktabs}
\usepackage[T1]{fontenc}
\usepackage{inputenc}  
\usepackage{esint}
\usepackage[authoryear]{natbib}
\usepackage{sidecap}
\usepackage[font=footnotesize]{caption}
\usepackage{pgfplots}
\usepackage{bbm}
\pgfplotsset{width=7cm,compat=1.8}

\usepackage{float}

\hypersetup{pdfpagemode=None, colorlinks=true,
 anchorcolor= webbrown, citecolor= blue,
filecolor= webbrown, linkcolor= blue, menucolor= webbrown,
urlcolor= webbrown, citebordercolor= 1 0 0, menubordercolor=1 0
0, urlbordercolor=1 0 0, runbordercolor=1
0 0 } \hypersetup{pdfauthor=Bruno Levy}
\definecolor{webgreen}{rgb}{.6,.6,.6}
\definecolor{webbrown}{rgb}{.6,0.15,0.15}
\definecolor{webyellow}{rgb}{0.98,0.92,0.73}

\newcommand\ddate{\today}
\newcommand\aauthorA{\footnote{brunopcl@al.insper.edu.br} Bruno P. C. Levy  }
\newcommand\aauthorB{\footnote{hedibertfl@insper.edu.br} Hedibert F. Lopes}
\newcommand\aaffiliationA{Insper}
\newcommand\aaffiliationB{Insper}
\newcommand\eemailA{\href{mailto:brunopcl@al.insper.edu.br}{{\small\texttt{{}}}}}
\newcommand\eemailB{\href{mailto:hedibertfl@insper.edu.br}{{\small\texttt{{}}}}}
\newcommand\ttitle{Dynamic Portfolio Allocation in High Dimensions using Sparse Risk Factors}
\newcommand\tthanks{We thank to all participants at Insper Seminars for useful comments. All remaining errors are of our responsibility.
}

\newcommand{\propref}[1]{\hyperref[#1]{Proposition~\ref*{#1}}}
\newcommand{\appref}[1]{\hyperref[#1]{Appendix~\ref*{#1}}}
\newcommand{\secref}[1]{\hyperref[#1]{Section~\ref*{#1}}}

\newcommand{\commentout}[1]{}

\newtheorem*{lemma*}{Lemma}

\hypersetup{pdfstartview=FitH}
\title{\vspace{-1.5\baselineskip}  
{\ttitle\thanks{~\tthanks}}\bigskip}

\author{{\aauthorA} \\   
{\small \aaffiliationA}\\
[-.1in] \eemailA \and
{\aauthorB} \\       
{\small \aaffiliationB}\\
[-.1in] \eemailB \medskip }
\date{\normalsize \textcolor{webgreen}{\ddate}}
\date{\normalsize \textcolor{webgreen}{Working Paper - November, 2021}}

\begin{document}

\maketitle

\vspace{-1.5\baselineskip}

\begin{abstract} 

We propose a fast and flexible method to scale multivariate return volatility predictions up to high-dimensions using a dynamic risk factor model.
Our approach increases parsimony via time-varying sparsity on factor loadings and is able to sequentially learn the use of constant or time-varying parameters and volatilities. We show in a dynamic portfolio allocation problem with 452 stocks from the $S\&P$ 500 index that our dynamic risk factor model is able to produce more stable and sparse predictions, achieving not just considerable portfolio performance improvements but also higher utility gains for the mean-variance investor compared to the traditional Wishart benchmark and the passive investment on the market index.

\vspace{1cm}


\vspace{1cm}

\noindent\noindent \emph{J.E.L. codes: C32, C52, C53, C58, G11  } 

\end{abstract}

\onehalfspace

\newpage

\tableofcontents

\newpage

\section{Introduction} 
\label{sec1: Introduction}

Portfolio allocation is one of the most common problems in finance. Since the seminal work of \cite{markowitz1952portfolio}, mean-variance optimization has been the most traditional way to select stocks in the financial industry and is commonly applied in the academic literature. The covariance matrix of returns is the key input to generate optimal portfolio weights, which makes its forecast accuracy crucial for out-of-sample portfolio performance. However, the universe of assets available for allocation is vast nowadays, increasing the dimension of such covariance matrices potentially to hundreds or even thousands of stocks. Due to the fact that the number of parameters in a covariance matrix grows quadratically with the number of assets, it inserts the traditional  Markowitz's problem into the curse of dimensionality.

Since large covariance matrices are extremely susceptible to estimation errors and instabilities, producing poor out-of-sample predictions for portfolio construction,  we propose what we call a \textit{Dynamic Risk Factor Dependency Model} (DRFDM). By the use of economically motivated risk factors and inducing time-varying sparsity on factor loadings, our approach is able to achieve higher model parsimony, dramatically reducing the parameter space and improving predictions for final portfolio decisions. The DRFDM combines a factor structure with sparsity in a conjugate and sequential fashion without the use of MCMC schemes, making the estimation process much faster and allowing investors to backtest a universe of hundreds of assets in a matter of few minutes.

The use of factor models in the financial literature is not new. With different applications in the asset pricing literature, since the CAPM of \cite{sharpe1964capital}, the APT of \cite{ross1976arbitrage} and the seminal work of \cite{fama1992cross}, these models consider that all systematic variation in returns are driven by a set of common factors that can be observable financial indices or unknown latent variables. This framework is commonly used for evaluating return anomalies and portfolio manager performance but also for portfolios construction. Our paper will focus on the use of observable risk factor models to build optimal portfolios.  

To illustrate the idea behind factor models, consider an exact K-risk factor model with a N-dimensional vector of assets returns:

\begin{equation}
\boldsymbol{r}_{t}= \boldsymbol{\alpha}_{t} + \boldsymbol{B}_{t} \boldsymbol{f}_{t} 
+ \boldsymbol{\epsilon_{t}},
\quad \quad \boldsymbol{\epsilon_{t}} \mid \boldsymbol{\Omega}_{t} \sim \mathcal{N}\left(\boldsymbol{0}, \boldsymbol{\Omega}_{t}\right)
\label{eq_1}
\end{equation}

\noindent where $\boldsymbol{B}_{t}$ is a N $\times$ K matrix of factor exposures (loadings) to the K risk factors $\boldsymbol{f}_{t} $ and $\boldsymbol{\Omega}_{t} = \operatorname{diag}
\left(\sigma_{1 t}^{2}, \ldots, \sigma_{N t}^{2}\right)$. If $Var(\boldsymbol{f}_{t} ) = \boldsymbol{\Sigma_{t}^{f}}$, the model in (\ref{eq_1}) implies an unconditional variance for return as:

\begin{equation}
Var(\boldsymbol{r}_{t}) =    \boldsymbol{\Sigma}_{t}^{r}=  \boldsymbol{B}_{t} \boldsymbol{\Sigma}_{t}^{f} \boldsymbol{B}_{t}^{\prime} + \boldsymbol{\Omega}_{t}
\label{eq_2}
\end{equation}

\vspace{3mm}

\noindent which is divided between a systematic component related to the factor exposures and factor covariances and  an idiosyncratic component for each individual asset return. Note that using K $<<$ N implies a strong reduction on the total number of parameters to be estimated in $\boldsymbol{\Sigma}_{t}^{r}$. 

When latent factor models are considered, $\boldsymbol{f}_{t} $ is treated as an unobserved variable estimated from the data. Some important references are \cite{aguilar2000bayesian}, \cite{han2006asset}, \cite{lopes2007factor}, \cite*{carvalho2011dynamic}, \cite*{zhou2014bayesian}, \cite*{kastner2017efficient} and \cite{kastner2019sparse}. Both \cite{carvalho2011dynamic} and \cite{zhou2014bayesian} explore the notion of time-varying sparsity, with factor loadings being equal to zero for different periods of time. \cite{kastner2019sparse} induces static sparsity by the use of shrinkage priors on factor loadings, pulling coefficients toward zero. Although his approach induces greater parsimony, shrinkage priors do not impose coefficients to be exactly zero, remaining a portion of estimation uncertainty that is still carried to the covariance matrix. 

The major drawback of latent factor models is that it requires MCMC schemes to simulate from joint posteriors, imposing great computational burden to scale up models to high-dimensions. The problem is itensified by the fact that strategies in quantitative finance require a sequential analysis for backtests, so the MCMC needs to be repeated for each period of time, which can take not just several days but weeks to be completed. Therefore, latent factor models are prohibitive for sequential analysis in high-dimensions (see \citealp{gruber2016gpu, gruber2017bayesian} and \citealp{west2020bayesian} for a deeper discussion on model scalability).

In order to seek for fast sequential analysis and higher model flexibility, instead of estimating latent factors, we include observable risk factor commonly used in the financial literature to represent the common movements of returns. We use the traditional 5 factors from \cite{fama2015five} as the main representation, also showing results for different subsets of risk factors. There are different papers in the literature that have already addressed the estimation of the covariance matrix of returns using observable risk factors. Recent key references include \cite*{wang2011dynamic}, \cite*{brito2018forecasting}, \cite*{puelz2020portfolio} and \cite*{de2018factor}. The main advantage of DRFDM in relation to those papers is its ability to take into account model uncertainty in a dynamic fashion in different model settings for individual asset returns. As we explain in the next section, the DRFDM uses what started to be known in the recent econometric literature as a "Decouple/Recouple" concept. The basic idea is to decouple the multivariate dynamic model into several univariate customized dynamic linear models (DLM) that can be solved in parallel and then be recouple for forecasting and decisions. It is strictly related to the popular Cholesky-style Multivariate Stochastic Volatility of \cite*{lopes2016parsimony}, \cite*{shirota2017cholesky} and \cite{primiceri2005time} and also applied in a similar fashion in \cite*{zhao2016dynamic}, \cite*{fisher2020optimal}, \cite*{lavine2020adaptive} and \cite{levy2021dynamic}. This framework allow us to take the model uncertainty problem into the univariate context, making highly flexible dynamic model choices.

Since it is well known that the environment of the economy and the financial market is continuously changing, model flexibility becomes a  extremely appealing feature to be incorporated nowadays. The common patterns in stock returns in the 90s are different from those during the Great Financial Crisis or during the recent Covid-19 pandemic. Factor exposures can be lower or higher, depending on calm or stressed periods. Additionaly, we can consider the fact that for some periods a subset of stock returns are not loading on specific risk factors. This motivates our work to impose time-varying sparsity, where inspired by the works of \cite*{raftery2010online}, \cite{dangl2012predictive} and \cite{koop2013large} we use a \textit{Dynamic Model  Selection} (DMS) approach to sequentially select  risk factors via dynamic model probabilities, where a risk factor is included if it is empirically wanted. We also consider a model space that is determined not just by different risk factors, but also by different degrees of variation in factor loadings, return volatilities and factor volatilities. Therefore, the DRFDM can adapt to environments of higher, lower or no variation in coefficients, for each specific entry of matrices in Equation (\ref{eq_2}). Hence, one specific asset return can be much more volatile than others and some risk factors can vary differently over time, also moving from constant to time-varying parameters. This last setting has a similar flavour of parsimony in the sense of \cite{lopes2016parsimony}.

Dynamic risk factor selection also offers an important benefit in terms of model parsimony since it imposes factor loadings of non-selected factors to be exactly equal to zero. It tends to induce much higher parsimony compared to continuous shrinkage priors applied in \cite{kastner2019sparse} and many other papers nowadays. As argued in \cite*{huber2020inducing} and \cite*{hauzenberger2020combining}, continuous shrinkage priors offer a lower bound of accuracy to be achieved and, for highly parametrized models, parameter uncertainty over-inflate predictive variances. 

Finally, an additional contribution of DRFDM to the literature relies on its simple  and fast computation. Since we rely on a conjugate model, with closed-form solutions for posterior distributions and predictive densities, there is no need for expensive MCMC methods, making the whole process extremely fast and easily scalable for high-dimensions. As we show in the empirical section, we perform a portfolio allocation procedure using data covering both the Great Financial Crisis and the initial stage of the Covid-19 pandemic with almost 500 stock returns from the $S\&P$ 500 index that can be backtested in just few minutes. We compare results with different specification choices and the traditional Wishart Dynamic Linear Model (W-DLM) as a benchmark. The W-DLM has been a standard model in the Bayesian financial time series and in the financial industry, because of its scalability and availability of sequential filtering. Our results show that the DRFDM is able to produce not just much stronger statistical improvements but substantial increase in Sharpe Ratios and risk reduction. We also show that a mean-variance investor will be willing to pay a considerable management fee to switch from the W-DLM  benchmark, from different well known estimation methods in the literature and the passive investment on the $S\&P$ index to the \textit{Dynamic Risk Factor Dependency Model}. 
 
The remainder of the paper is organized as follows.  The general econometric framework is introduced in Section \ref{sec2: Econometric}. Section \ref{sec3: Factor Selection} details the  time-varying model and factor selection approach and how it can be applied to impose sparsity. In Section \ref{sec4: Empirical} we perform our empirical analysis, providing an out-of-sample statistical and economic performance evaluation in a high-dimensional environment. Section \ref{sec5: Conclusion} concludes.

 \vspace{4mm}

\section{Econometric Framework} \label{sec2: Econometric}

As mentioned in Section \ref{sec1: Introduction}, our work is inspired by the \textit{Cholesky-style} framework in \cite{lopes2016parsimony} and \cite{primiceri2005time}, being closely related to the Dynamic Dependency Network Model of \cite{zhao2016dynamic}. The great advantage of this framework is to model the cross-sectional contemporaneous relations among different series and customize univariate DLMs. Consider $\boldsymbol{r_{t}}$ as a $\textit{N}$-dimensional vector with asset returns time series $r_{j,t}$ and consider the following dynamic system:


\begin{equation}
\left(\boldsymbol{I}_N-\boldsymbol{B}_{t}\right)
\boldsymbol{r}_{t}= \boldsymbol{\alpha}_{t}
+ \boldsymbol{\epsilon}_{t},
\quad \quad \boldsymbol{\epsilon}_{t} \mid \boldsymbol{\Omega}_{t} \sim \mathcal{N}\left(\boldsymbol{0}, \boldsymbol{\Omega}_{t}\right),
\label{eq_3}
\end{equation}

\noindent where $\boldsymbol{\alpha}_{t}$ is a $\textit{N}$-dimensional vector of time-varying intercepts and  $\boldsymbol{\Omega}_{t} = \operatorname{diag}
\left(\sigma_{1 t}^{2}, \ldots, \sigma_{N t}^{2}\right)$.
All contemporaneous relations among different asset returns are coming from the $\textit{N}$ $\times$ $\textit{N}$ matrix $\boldsymbol{B}_{t}$, whose off-diagonal elements $\beta_{ji t}$s (for $j \neq i $) capture the dynamic contemporaneous relationships among series $j$ and $i$ at time $t$ and
$\boldsymbol{B}_{t}$ has zeroes on the main diagonal.

In the work of \cite{lopes2016parsimony}, \cite{zhao2016dynamic} and \cite{levy2021dynamic}, $\boldsymbol{B}_{t}$ is a lower triangular matrix with zeroes in and above the main diagonal:

\vspace{3mm}

\begin{equation}
\boldsymbol{B}_t=\left[\begin{array}{ccccc}
0 & 0 & \ldots &\ldots &  0\\
\beta_{21, t} & 0 & \ldots & \ldots &  0\\
\vdots & \vdots & \ddots & \vdots & \vdots \\
\beta_{N 1, t} & \beta_{N 2, t} & \ldots & \beta_{N, N-1, t} & 0
\end{array}\right]
\label{eq:triangular}
\end{equation}

\vspace{3mm}

Since the error terms in  $\boldsymbol{\epsilon_{t}}$ are contemporaneouly uncorrelated, the triangular contemporaneous dependencies among asset returns in Equation (\ref{eq:triangular}) generate a fully recursive system, known as a Cholesky-style framework (\citealp{west2020bayesian}). Hence, each equation $j$ of the system will have its own set of $\textit{parents}$ ($\boldsymbol{r_{pa(j), t}}$), that is, will depend contemporaneously on all other asset returns above equation $j$, following the triangular format in Equation (\ref{eq:triangular}). In words, the top asset return in the system will not have parents, the second from the top asset return will have the first time series of returns as a parent and will load on it, the third asset return will have the first two asset returns as parents and will load on them all the way to the last asset return, which will depend on all other $N-1$ returns above it.

Equation (\ref{eq_3}) can be rewritten in the reduced form as

\begin{equation}
\boldsymbol{r}_{t}=\boldsymbol{A}_t \boldsymbol{\alpha}_{t} + \boldsymbol{u}_{t}, \quad    \quad \boldsymbol{u}_{t} \mid \boldsymbol{\Sigma}_{t} \sim \mathcal{N}\left(0, \boldsymbol{\Sigma}_{t}\right)
\label{eq:reducedform}
\end{equation}

\noindent where $\boldsymbol{A}_t=\left(\boldsymbol{I}_N-\boldsymbol{B}_{t}\right)^{-1}$ and $\boldsymbol{u}_t = \boldsymbol{A}_t \boldsymbol{\epsilon}_{t}$. The modified Cholesky decomposition clearly appears in $\boldsymbol{\Sigma}_{t}=\boldsymbol{A}_t \boldsymbol{\Omega}_{t}\boldsymbol{A}_t^{\prime}$ which is now a full variance-covariance matrix capturing the contemporaneous relations among the $\textit{N}$ asset returns. Given the parental triangular structure of $\boldsymbol{B_{t} }$ in (\ref{eq:triangular}), the equations will be conditionally independent, bringing the ``Decoupled'' aspect of the multivariate model. In other words, the multivariate model can be viewed as a set of $\textit{N}$ conditionally independent univariate DLMs that can be dealt with in a parallelizable fashion. The outputs of each equation are then used to compute $\boldsymbol{B_{t}}$ and $\boldsymbol{\Omega_{t}}$, hence recovering the full time-varying covariance matrix $\boldsymbol{\Sigma_{t}}$.

\subsection{Dynamic Risk Factor Dependency Model}

Although the model above shows greater flexibility, it remains highly parameterized and susceptible to producing poor out-of-sample forecasts. Notice that for a model with hundreds or thousands of equations, those asset returns at the bottom part will load on several hundreds of other assets, making out-of-sample forecasts too unstable. 

Additionally, the triangular form in Equation (\ref{eq:triangular}) makes the system dependent of the asset return ordering. As highlighted by \cite{levy2021dynamic}, it is an important drawback of the \textit{Cholesky-style} framework, because imposing a specific order structure can lead to inferior final decisions and  harm portfolio performance. Since the "correct" series ordering is uncertain and the environment of the economy is continuously changing,  \cite{levy2021dynamic} propose what they call a \textit{Dynamic Ordering Learning}. That is a flexible model that deals with the ordering uncertainty in a dynamic fashion, where the econometrician is able to sequentially learn the contemporaneous relations among different series over time. However, since there are $N$! possible orders to learn for each period of time, even with a flexible model it is a prohibitive task in high-dimensions.   

In order to impose greater model parsimony and overcome the ordering uncertainty in high-dimensions, we propose what we call a \textit{Dynamic Risk Factor Dependency Model}. Inspired by the literature on observable risk factors, in this method we augment the $N$-vector of asset returns with $K$ economically motivated risk factors, for $K << N$,  and impose the restriction that all dynamic contemporaneous dependencies among asset returns are coming from them. It drastically reduces both the parameter space and the ordering uncertainty problem to a low-dimension one, being easily implemented by the \textit{Dynamic Ordering Learning} approach of \cite{levy2021dynamic}.

Defining a new vector of returns $\boldsymbol{\mathcal{R}}_{t} = \left(\boldsymbol{\mathcal{F}}_{t}\quad , \quad \boldsymbol{r}_{t} \right)^{\prime} $ , augmented by the $K$-dimensional vector of known risk factors, $\boldsymbol{\mathcal{F}}_{t}$, we rewrite Equation (\ref{eq_3}) as:

\vspace{4mm}

\begin{equation}
\left(\boldsymbol{I}_{K+N}-\boldsymbol{\mathcal{B}}_{t}\right)
\boldsymbol{\mathcal{R}}_{t}= \boldsymbol{\Tilde{\alpha}}_{t}
+ \boldsymbol{\Tilde{\epsilon}}_{t},
\quad \quad \boldsymbol{\Tilde{\epsilon}}_{t} \mid \boldsymbol{\Tilde{\Omega}}_{t} \sim \mathcal{N}\left(\boldsymbol{0}, \boldsymbol{\Tilde{\Omega}}_{t}\right),
\label{eq:ddnm2}
\end{equation}

\vspace{4mm}

\noindent where the tilde upscript represents the extended to $K+N$ dimension version of previous vectors and matrices of Equation (\ref{eq_3}). Now, $\boldsymbol{\mathcal{B}}_{t}$ is a new $(K+N) \times (K+N)$ matrix containing all the dynamic contemporaneous dependencies, where both factor and asset returns are allowed to load only on the set of observable chosen factors:

\begin{equation}
\small
\boldsymbol{\mathcal{B}}_{t}=\left[\begin{array}{ccccccccc}
 0& 0 & \ldots & \ldots &  \ldots & \ldots &  \ldots &  \ldots &0 \\
\beta_{2,1, t} & 0 & \ldots & \ldots &  \ldots & \ldots &  \ldots & \ldots &0 \\
\beta_{3,1, t} & \beta_{3,2, t} & 0 & \ldots &  \ldots & \ldots &  \ldots &  \ldots &0 \\
\vdots & \vdots & \ddots &  \ddots &  \vdots &   \vdots &   \vdots &  \vdots &  \vdots\\
\beta_{K, 1, t} & \beta_{K, 2, t} & \ldots & \beta_{K, K-1, t} &  0  &  0 &  \ldots &   \ldots & 0\\
\beta_{K+1, 1, t} & \beta_{K+1, 2, t} & \ldots & \beta_{K+1, K-1, t} & \beta_{K+1, K, t}  &  0 &  \ldots &\ldots &0    \\
\beta_{K+2, 1, t} & \beta_{K+2, 2, t} & \ldots & \beta_{K+2, K-1, t} & \beta_{K+2, K, t} & 0   &  \ldots & \ldots &0    \\

\vdots & \vdots & \ldots & \vdots & \vdots & \vdots  &  \ldots & \ldots & 0 \\
\beta_{K+N, 1, t} & \beta_{K+N, 2, t} & \ldots & \beta_{K+N, K-1, t} & \beta_{K+N, K, t} &  0  &  \ldots &  \ldots & 0  \\

\end{array}\right]
\label{eq:triangular2}
\end{equation}

\vspace{4mm}

Our approach can be viewed as an extension of  the work of \cite{zhao2016dynamic}, however, instead of following the whole triangular format as in Equation (\ref{eq:triangular}), we impose that any asset return dependencies are coming from common factors in the economy and not on specific movements of different asset returns. Hence, for $j = 1,\dots,K$, the matrix $\boldsymbol{\mathcal{B}}_{t}$ follows the usual triangular format. However, for $j > K$, asset returns are restricted to load until series (factor) $K$. This new representation allows us to estimate a much lower number of parameters, since  $\boldsymbol{\mathcal{B}}_{t}$ is filled with zeroes after the $K$-th column. The sparser $\boldsymbol{\mathcal{B}}_{t}$ is, the more stable and efficient the resulting inferences are, producing better out-of-sample predictions for decision analysis. Additionaly, it gives higher economic intuition for the variation of asset returns, since the common movements follow exposures to well know risk factors developed by the financial literature.

\subsection{$\textit{K + N}$ univariate dynamic linear models}

As mentioned before, given the structure of $\boldsymbol{\mathcal{B}}_{t}$, the set of $K+N$ univariate models can be represented as $K+N$ univariate recursive dynamic regressions, where we have for each $j = 1, \dots, K, \dots, K+N$: \footnote{Inspired on recent evidence of momentum on risk factors (\citealp{gupta2019factor}) we also have tested risk factors as a AR(1) process, but results were very similar so we decided to maintain in this paper the traditional structure of Equation (\ref{eq:dlms1}). }

\begin{equation}
\mathcal{R}_{j t} =\alpha_{j, t}  +\boldsymbol{\mathcal{R}}_{pa(j), t}^{\prime} \boldsymbol{\beta}_{pa(j), t}+\nu_{j t},  \quad \quad \nu_{j t} \sim \mathcal{N}\left(0, \sigma_{j t}^{2}\right),
\label{eq:dlms1}
\end{equation}

\noindent where for $j = 1, \dots, K$, the \textit{parental set} $\boldsymbol{\mathcal{R}}_{pa(j), t} = \boldsymbol{\mathcal{F}}_{pa(j), t}$ represents all risk factor series in $\boldsymbol{\mathcal{R}}_{t}$ that are above series $j$ and, for $j > K$, $\boldsymbol{\mathcal{R}}_{pa(j), t}$ possibly represents all series until series K, i.e, all risk factors in $\boldsymbol{\mathcal{F}}_{t}$. 

We represent the dynamic coefficients in  Equation (\ref{eq:dlms1}) evolving according to random walks:

\begin{equation}
\left(\begin{array}{c}
\alpha_{j, t} \\
\boldsymbol{\beta}_{pa(j), t}
\end{array}\right)=\left(\begin{array}{c}
\alpha_{j, t-1} \\
\boldsymbol{\beta}_{pa(j),  t-1}
\end{array}\right)+\boldsymbol{\omega}_{j t} \quad \boldsymbol{\omega}_{j t} \sim \mathcal{N}\left(\boldsymbol{0}, \boldsymbol{W}_{j t}\right).
\end{equation}

\vspace{3mm}

\noindent and by ease of exposition, we define $x_{j, t-1}=1 $ and represent the full dynamic state and regression vectors as

$$
\boldsymbol{\theta}_{j t}=\left(\begin{array}{c}\alpha_{j, t} \\ \boldsymbol{\beta}_{pa(j), t}\end{array}\right) \quad \text{and} \quad \mathbf{F}_{j t}=\left(\begin{array}{c} x_{j, t-1} \\ \boldsymbol{\mathcal{R}}_{pa(j), t}\end{array}\right),
$$

\vspace{3mm}

Hence, denoting $y_{jt}$ as the $j$-$th$ variable in $\boldsymbol{\mathcal{R}}_{t}$ we recover the traditional univariate DLM formulation as in \cite{west2006bayesian}, namely

\begin{eqnarray*}
y_{j t} &=& \mathbf{F}_{j t}^{\prime} \boldsymbol{\theta}_{j t}+\nu_{j t},  \quad \quad \nu_{j t} \sim N\left(0, \sigma_{j t}^{2}\right),\\
\boldsymbol{\theta}_{j t} &=& \boldsymbol{\theta}_{j,t-1}+\boldsymbol{\omega}_{j t},  \quad \quad \boldsymbol{\omega}_{j t} \sim N\left(\boldsymbol{0}, \boldsymbol{W}_{j t}\right),
\end{eqnarray*}

\noindent for $j=1,\ldots, K + N$, where again $\boldsymbol{\theta}_{j t}$ evolves over time as a simple random-walk. 


\paragraph{Posterior at $t-1$.}  Following the algorithmic structure of sequential learning in DLMs \cite{west2006bayesian}, Chapter 4, at time $t-1$ and for each time series $j$, the joint posterior distribution of $\boldsymbol{\theta}_{j t-1} $ and $\sigma_{j t-1}$ at time $t-1$ is a multivariate Normal-Gamma\footnote{In our empirical analysis we have considered initial states centering $\theta_{j0}$ around zero ($m_{j, 0}=\boldsymbol{0}$) with $C_{j,0} = 100\boldsymbol{I}$. We set $s_{j, 0}$ as the sample variance estimate of residuals from an OLS model over the training period and we start with $n_{j,0} = 10$. }:

\begin{equation}
\boldsymbol{\theta}_{j, t-1}, \sigma_{j t-1}^{-1} \mid \mathcal{D}_{t-1} \sim \mathcal{NG}\left(\boldsymbol{m}_{j, t-1}, \boldsymbol{C}_{j, t-1}, n_{j, t-1}, n_{j, t-1} s_{j, t-1}\right).
\label{eq:postt1}
\end{equation}


\noindent 
Through the random walk evolution and conjugacy, we can derive the joint prior distribution of $\boldsymbol{\theta}_{jt} $ and $\sigma_{jt}$ for time $t$ as: 
\begin{equation}
\boldsymbol{\theta}_{jt}, \sigma_{jt}^{-1} \mid \mathcal{D}_{t-1} \sim\mathcal{NG}\left(\boldsymbol{a}_{j t}, \boldsymbol{R}_{j t}, r_{j t}, r_{j t} s_{j, t-1}\right)
\end{equation}
where $r_{j t}=\kappa_{j} n_{j, t-1}$, $\boldsymbol{a}_{j t} = \boldsymbol{m}_{j, t-1}$ and  $\boldsymbol{R}_{j t} = \boldsymbol{C}_{j, t-1} / \delta_{j}$.  The quantities $0 < \delta_{j} \leq 1$ and $0  < \kappa_{j} \leq 1$ represent specific discount (aka forgetting) factors for $\theta_{jt}$ and $\sigma_{jt}$, respectively. Discount methods are used to induce time-variations in the evolution of parameters and have been extensively used in many applications (\citealp{raftery2010online}, \citealp{dangl2012predictive}, \citealp{koop2013large}, \citealp*{mcalinn2020multivariate}, amongst others) and well documented in 
\cite{west2006bayesian}, \cite{lopes2006mcmc} and \cite{prado2010time}.

\paragraph{1-step ahead forecast at $t-1$.} The (prior) predictive distribution of $y_{j t}$ is a Student's $t$ distribution with $r_{j t}$ degrees of freedom:

$$
y_{j t} \mid \boldsymbol{y}_{pa(j),t}, \mathcal{D}_{t-1} \sim \mathcal{T}_{r_{j t}}\left(f_{j t} ,   q_{j t}\right),
$$

\noindent with $f_{j t} = \mathbf{F}_{j t}^{\prime}\boldsymbol{a_{j t}}$ and $q_{j t} = s_{j, t-1}+\mathbf{F}_{j t}^{\prime} \mathbf{R}_{j t} \mathbf{F}_{j t} $. It is important to notice that in this framework we have a conjugate analysis for forward filters and one-step ahead forecasting. Therefore, we are able to compute closed-form solution for predictive densities for each equation $j$. Hence, conditional on $\textit{parents}$, it is easy to compute the joint predictive density for $\boldsymbol{y_{t}} $:

\begin{equation}
p\left(\boldsymbol{y}_{t} \mid \mathcal{D}_{t-1}\right)=\prod_{j=1}^{(K + N)} p\left(y_{j t} \mid \boldsymbol{y}_{pa(j) t}, \mathcal{D}_{t-1}\right),
\label{eq:predictive}.
\end{equation}

\noindent which simply is the product of the already computed $K + N$ different univariate Student's $t$ distributions. After the time series are decoupled for sequential analysis, they are then recoupled for multivariate forecasting. In our decision analysis at Section \ref{sec4: Empirical}, we divide the recoupled part in two: one related to the dynamics of the $K$ factors and the other to the dynamic of the $N$ asset returns.  We will be concerned with the mean and variance of each of this parts for the portfolio allocation study:

\begin{equation}
\boldsymbol{\lambda}_{t|t-1}=E\left(\boldsymbol{\mathcal{F}}_{t} \mid \mathcal{D}_{t-1}\right) \ \ \ \mbox{and} \ \ \  \boldsymbol{\Sigma}_{t|t-1}^{f}=Var\left( \boldsymbol{\mathcal{F}}_{t} \mid \mathcal{D}_{t-1}\right).
\end{equation}

\noindent for the $K$-vector of expected factor means and the $K \times  K$ expected factor covariance matrix and

\begin{equation}
\boldsymbol{f}_{t|t-1}=E\left(\boldsymbol{r}_{t} \mid \mathcal{D}_{t-1}\right) \ \ \ \mbox{and} \ \ \  \boldsymbol{\Sigma}_{t|t-1}^{r}=Var\left(\boldsymbol{r}_{t} \mid \mathcal{D}_{t-1}\right).
\end{equation}

\noindent for the $N$-vector of expected asset returns and the  $N \times  N$ expected covariance matrix of returns. Further details about the derivations of the evolution, forecasting, updating distributions can be found in Appendices A and B.

\vspace{4mm}

\section{Dynamic Model and Factor Selection}
\label{sec3: Factor Selection}

As argued by \cite{kastner2019sparse}, even imposing parsimony by a factor structure, models are still rich in parameters when considering a high-dimension portfolio problem. It motivates our work to induce stronger parsimony by the use of sparsity on factor loadings in a time-varying and sequential form. This approach is convenient to improve out-of-sample predictions and introduce greater model flexibility, since for each period of time different assets can load just on a subset of risk factors. 

Model uncertainty is a well-known challenge among applied researchers and industry
practitioners that are interested in producing forecasts for decision-making problems. In the last decades, the Bayesian literature has addressed this question with great success. Bayesian Model Averaging and Bayesian Model Selection are well-known methodologies for static models when there is uncertainty
about the predictors to include (\citealp{madigan1994model} and \citealp{hoeting1999bayesian}). More recently,  \cite{raftery2010online} have propose a dynamic version of Bayesian Model Selection, called Dynamic Model Selection (DMS). They suggest the use of reasonable approximations, borrowing ideas from discount (forgetting) methods, avoiding simulations of transition probabilities matrices and maintaining the conjugate form of posterior distributions, allowing analytical solutions for forward filtering and forecasting which significantly reduces the computation burden of the process. Their approach has also shown great success, being applied recently in macroeconomics and finance (\citealp{dangl2012predictive}, \citealp{koop2013large}, \citealp{catania2019forecasting} and \citealp{levy2021dynamic}). 

In our portfolio allocation problem we apply DMS for each individual equation of the system in (\ref{eq:ddnm2}), learning sequentially about which specifications to choose, such as the main risk factors and the best discount factors conducting time-variation in factor loadings and volatilities. Therefore, not just risk factors can have a stochastic or constant volatility, but also each individual asset return can have constant or time-varying factor loadings and volatilities, switching its behavior depending on the environment of the economy. Using DMS we are able to dynamically select the best risk factors by imposing factor loadings equal to zero for non-selected  risk factors. For a highly parametrized model it can be viewed as an advantage compared to traditional shrinkage priors that set coefficients close but not equal to zero, increasing uncertainty on predictions.   

It is important to highlight here that the main focus of our dynamic factor selection approach is to induce model sparsity in order to deflate the covariance structure among the universe of asset returns and to produce more stable predictions of expected returns. Although we take advantage of the fact that our DRFDM is able to predict expected returns and we use this predictions for portfolio decisions, we will not be particularly concerned  on inferences about asset pricing models or best risk factors to explain stock returns. For recent advances on Bayesian model selection/averaging with focus on investigating expected returns based on factor asset pricing models, we refer to \cite*{bryzgalova2019bayesian} and \cite{hwang2020bayesian}.

\subsection{Dynamic Model Probabilities}
\label{sec3.1}

In order to perform DMS in each equation, we introduce here the idea of dynamic model probabilities. Consider the case of a specific equation $j$ in the system discussed on the previous section. Suppose the dependent serie of this equation can load on $p$ different risk factors available. Hence, there are $2^{p}$ possible combinations of models defined by the \textit{parental set}. Also, when considering $n_{\delta}$ and $n_{\kappa}$ different possible discount factors for factor loadings and volatilities, the equation $j$ will have a total of $n_{j} = 2^{p}\times n_{\delta}\times n_{\kappa} $ possible models to choose in the univariate model space.\footnote{In our portfolio study we actually consider $n_{j} = (2^{p}-1)\times n_{\delta}\times n_{\kappa} $ possible models, since we exclude the case where an asset return is not loading in any risk factor.} 

The DMS approach deals with the model uncertainty by assigning probabilities for each possible model. Denote $\pi_{t-1 | t-1, i,j} = p(\mathcal{M}_{i}^{j} \mid  \mathcal{D}_{t-1})$ as the posterior probability of model $i$ and equation $j$ at time $t-1$. Following \cite{raftery2010online}, the predicted probability of the model $i$ given all the data available until time $t-1$ is expressed as:

 \begin{align}
 \pi_{t | t-1, i, j}=\frac{\pi_{t-1 | t-1, i, j}^{\alpha}}{\sum_{l=1}^{M} \pi_{t-1 | t-1, l, j}^{\alpha}},
\end{align}

\noindent where $0 \leq \alpha \leq 1$ is a forgetting factor. The main advantage of using $\alpha$ is avoiding the computational burden associated with expensive MCMC schemes to simulate the transition matrix between possible models over time. This approach has also been extensively used in the Bayesian econometric literaure in the last decade (\citealp{koop2013large}, \citealp{zhao2016dynamic}, \citealp{lavine2020adaptive} and \citealp{beckmann2020exchange}). After observing new data at time $t$, we update our model probabilities following a simple Bayes’ update:

\begin{align}
 \pi_{t | t, i,j}=\frac{\pi_{t | t-1, i,j}p_{i}\left(y_{j t} \mid \boldsymbol{y}_{pa(j) }, \mathcal{D}_{t-1}\right)}{\sum_{l=1}^{M} \pi_{t | t-1, l}p_{l}\left(y_{j t} \mid \boldsymbol{y}_{pa(j) }, \mathcal{D}_{t-1}\right)}.
\end{align}

\noindent the posterior probability of model $i$ at time $t$ where $p_{i}(y_{j t} \mid \boldsymbol{y}_{pa(j)}, \mathcal{D}_{t-1})$ is the predictive density of model $i$ evaluated at $y_{jt}$. Hence, upon the arrival of a new data point, the investor is able to measure the performance for each univariate model $i$ and to assign higher probability for those models that generate better performance. 

One possible interpretation for the forgetting factor $\alpha$ is through its role to discount past performance. Combining the predicted and posterior probabilities, we can show that

\begin{equation}
\pi_{t \mid t-1, i, j} \propto \prod_{l=1}^{t-1}\left[p_{i}\left(y_{j t-l} \mid \boldsymbol{y}_{pa(j)}, \mathcal{D}_{t-l-1}\right)\right]^{\alpha^{l}}.
\label{eq:pis}
\end{equation}

Since $0 < \alpha \leq 1$, Equation (\ref{eq:pis}) can be viewed as a discounted predictive likelihood, where past performances are discounted more than recent ones. It implies that models that received higher performance in the recent past will produce higher predictive model probabilities. The recent past is controlled by $\alpha$, since a lower $\alpha$ discounts more heavily past data and generates a faster switching behavior between models over time.\footnote{The value of $\alpha$ is commonly selected to be very close to 1. Following the majority of the econometric literature, we set $\alpha = 0.99$ in our empirical study.}

The idea of DMS is to select the model with the highest model probability for each period of time. Given the fact that each equation is conditionally independent, as soon as we select the best model for each equation, the posterior model probability of the multivariate model can be viewed as a product of the $K+N$ univariate model probabilities: 

$$P(\mathcal{M}_{1:(K+N)}^{*}|\mathcal{D}_{t-1})  = \prod_{j=1}^{(K+N)} P(\mathcal{M}_{j}^{*}|\mathcal{D}_{t-1})  $$

\noindent where the asterisk symbol represents the selected model. A major benefit of our our approach is that each equation in the system in (\ref{eq:ddnm2}) is conditionally independent and model uncertainty can also be addressed independently for each equation $j$. Hence, if for each equation we have a possible model space of size $n_{j}$, it implies a total of $\sum_{j = 1}^{K+N} n_{j} $ possible models. In a high-dimension environment with hundreds of equations and hundreds of possible univariate models for each one, it is a massive reduction on the total model space, since in a multivariate model with no conditional independence would require a total of $\prod_{j = 1}^{K+N} n_{j} $ models.

The use of Dynamic Model Probabilities offers a substantial flexibility for our approach. The DMS approach allow us to customize individual series, proposing different models with different risk factors and variation in coefficients, switching between them in a dynamic fashion. It is a important advantage compared to the traditional Wishart-DLM, where all series are stuck to the same discount factors for state evolutions.

\subsection{Dynamic Factor Ordering Learning}
\label{sec3.2}

As highlighted in Section \ref{sec2: Econometric}, the \textit{Cholesky-style} framework is based on the triangular structure of Equation (\ref{eq:triangular}). Therefore, models within this framework will depend on the series ordering structure selected by the researcher, leading to different contemporaneous relations among series. The recent work of \cite{levy2021dynamic} show evidences of instabilities on the "correct" ordering structure and propose a method to dynamically deal with the problem of ordering uncertainty. By the use of Dynamic Ordering Probabilities, the authors propose a  \textit{Dynamic Ordering Learning} (DOL) approach where the econometrician is able to select or average the outputs of different orderings in a sequential fashion. They show that taking into account the ordering uncertainty among different series improves out-of-sample predictions and portfolio performance. 

In our high-dimension environment, the computations of the total number of possible orderings is prohibitive, since for $N$ financial time series, there are $N!$ possible orders to compute.\footnote{Just to give an idea of how fast the order space grows with the number of series, a multivariate model with 6 series has $6! = 720$ possible orders. Just increasing the model dimension to 10 series increases the number of possible orders to 3,628,800. It is easy to see that when we go to hundreds of series, the number of orders goes to infinity. } However, once we impose a factor dependency structure as in (\ref{eq:triangular2}), we restrict our ordering uncertainty to a low dimension, because just the top $K$ variables rely on a triangular format. Since $K<<N$, we are able to follow the DOL approach of \cite{levy2021dynamic}. The idea is to assign ordering probabilities for the risk factors in a similar manner as we discussed in Section (\ref{sec3.1}). However, instead of selecting the best factor ordering over time, we average the outputs of each possible order, weighting by their order probabilities. For more details of the DOL procedure we refer to the paper of \cite{levy2021dynamic}. 


\section{Empirical Analysis} \label{sec4: Empirical}

In order to test how the DRFDM performs in a real world problem, in this section we compare our method in terms of out-of-sample statistical and portfolio performance  compared to different model settings and the Wishart-DLM. Our data is based on weekly stock returns from the Standard \& Poor's 500 index. A stock is included if it was traded over the full horizon 2002-2020:5 and was a constituent of the index at some period during that time, resulting in $N = 452$ stocks over 959 time periods. The data was colected from Bloomberg Terminal and log-returns were used for the analysis. In our main empirical analysis,  we use as observable risk factors the 5 factors of \cite{fama2015five} downloaded for the same period from the website of Kenneth French.\footnote{The data is available at \url{https://mba.tuck.dartmouth.edu/pages/faculty/ken.french/data_library.html}. We transformed daily to weekly factor returns.} The five risk factors are the market excess return (MKT), a size factor (SMB), a value factor (HML), a profitability factor (RMW) and a investment factor (CMA). In our portfolio performance below, we also show results when considering different subset of risk factors, such as the first three factors (MKT, SMB and HM), the three factors plus the Momentum factor (\citealp{carhart1997persistence}) and the five factors plus Momentum.

\begin{table}[!htbp] \centering 
\small
  \caption{GICS sectors and the number of members within the data set} 
  \label{sectors} 
\begin{tabular}{@{\extracolsep{5pt}}l cc} 
\\[-1.8ex]\hline 
\hline \\[-1.8ex] 
 GICS sectors & Numer of members \\ 
\hline \\[-1.8ex] 

Consumer Discretionary & $63$ \\ 
Consumer Staples & $28$ \\ 
Energy & $30$ \\ 
Financials & $57$ \\ 
Health Care & $59$ \\ 
Industrials & $65$ \\ 
Information Technology & $49$ \\ 
Materials & $29$ \\ 
Real Estate & $29$ \\ 
Communication Services & $17$ \\ 
Utilities & $26$ \\ 
\hline \\[-1.8ex] 
\end{tabular} 
\end{table}

We group stocks based on eleven sectors  of the Global Industry Classification Standard (GICS). It makes easier the visual interpretation of figures below. Table \ref{sectors} list the number of stocks in each sector.

\begin{figure}[!h]
  \centering
  
  \begin{minipage}[b]{0.32\textwidth}
    \includegraphics[width=\textwidth]{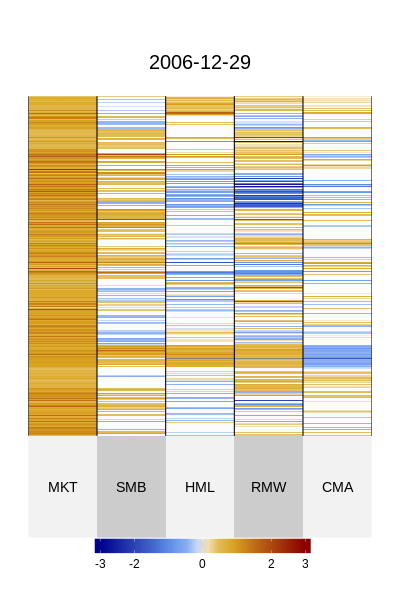}
  \end{minipage}
  \begin{minipage}[b]{0.32\textwidth}
    \includegraphics[width=\textwidth]{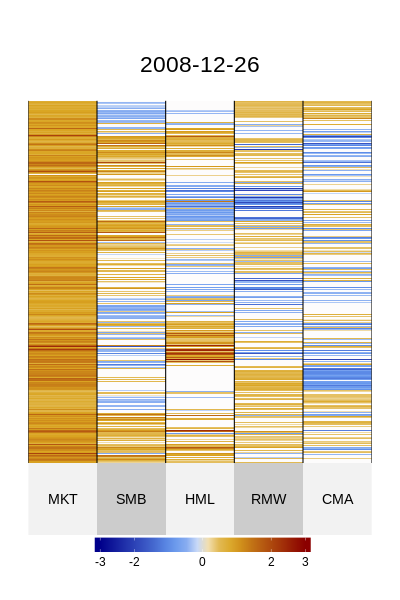}
  \end{minipage}
   \begin{minipage}[b]{0.32\textwidth}
    \includegraphics[width=\textwidth]{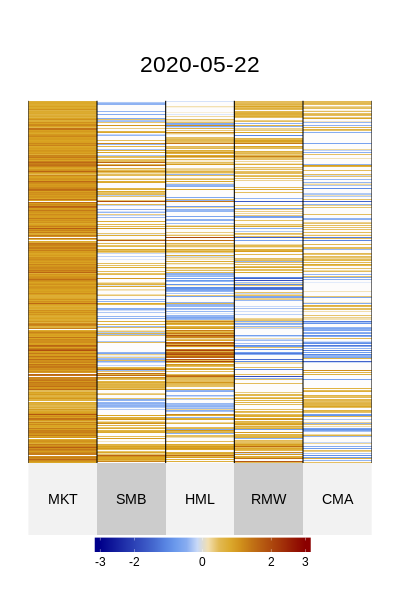}
  \end{minipage}
   \caption{\textit{Time-varying posterior mean of factor loadings for three different time periods. High positive factor loading are represented in dark red, strong negative factor loadings are in dark blue and zero factor loadings in white.}}
   \label{fig: factor_loadings}
\end{figure}

First, we illustrate in Figure \ref{fig: factor_loadings} the mean posterior factor loadings of our DRFDM with the five factors for three specific time periods. Each row represents the factor loads of an individual stock return loading on different risk factors of each column. The first figure at the left is referred to a calm time period, while the second and third dates were intentionally chosen to consider highly turbulence periods in the stock market, the Global Financial Crisis of 2008 and the the Covid-19 pandemic. The white colour represents factor loadings equal to zero. For the three periods, there is considerable sparsity on factor loadings, being stronger at the end of 2006 and weaker for the other stressed periods. Also note that not just the sparsity is changing over time, but also the mean of factor loadings are substantially varying, with periods of higher and lower values, with signs of many asset returns changing for the same risk factor. One interesting aspect of Figure \ref{fig: factor_loadings} is the great importance of the market factor for asset returns, being quite rare to observe sparsity on its loadings.

To clarify the dynamic sparsity pattern of our approach, Figure \ref{tv_loadings} in Appendix C focus on the time-varying movements of posterior mean factor loadings of two specific stocks of different sectors over time. In fact, Figure \ref{tv_loadings} highlights how differently they load on risk factors, with one company inducing much more sparsity than the other. Interestingly, with the exception of the market factor, all factor loadings are set to zero at least once. Also, depending on the company, some factor loadings are set to zero for the majority of the sample period.

\begin{figure}[!h]
  \centering
  
  \begin{minipage}[b]{0.32\textwidth}
    \includegraphics[width=\textwidth]{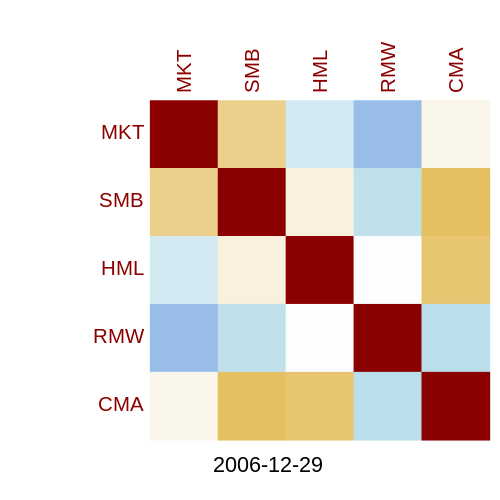}
  \end{minipage}
  \begin{minipage}[b]{0.32\textwidth}
    \includegraphics[width=\textwidth]{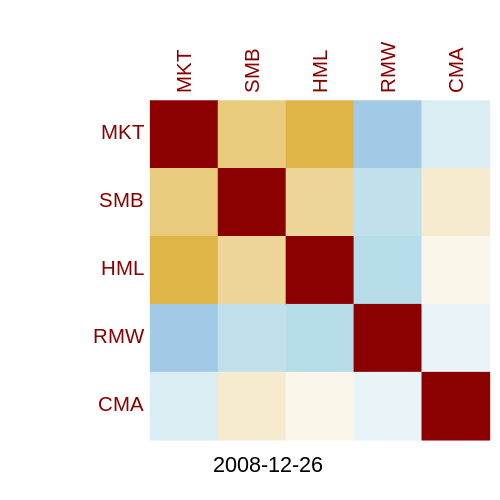}
  \end{minipage}
   \begin{minipage}[b]{0.32\textwidth}
    \includegraphics[width=\textwidth]{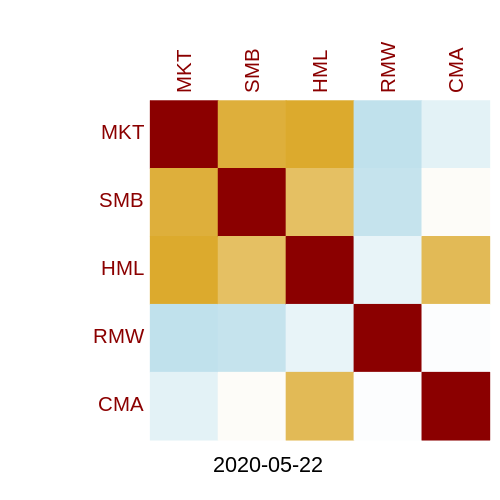}
  \end{minipage}
   \caption{\textit{Time-Varying predicted mean Factor Correlations. Colors scale follows the same pattern from Figure \ref{fig: factor_loadings}, but restricted to the \{-1, 1\} interval. Therefore, correlations equal to one are represented in dark red, negative perfect correlations in dark blue and zero correlations in white.} }
   \label{factors_cov}
\end{figure}

Although the estimation of factor loadings plays an important role, they are not the only ingredients to predict the covariance matrix of returns. As discussed in previous sections, the investor also needs to learn the time-variation on the factor covariance matrix. Figure \ref{factors_cov} shows the predicted mean of the time-varying factor correlation matrix over the same selected three periods. The first aspect to note is how the correlations among factors change over time. As an example is the sign changing in the correlation between the MKT and the HML factors at the end of 2006 to the stressed periods. Additionally, it seems that during bad periods the correlations tend to be stronger, specially for the three factors MKT, SMB and HML. One exception is the CMA factor, presenting very low factor correlations during these bad times.

\begin{figure}[!h]
  \centering
 
  \begin{minipage}[b]{0.32\textwidth}
    \includegraphics[width=\textwidth]{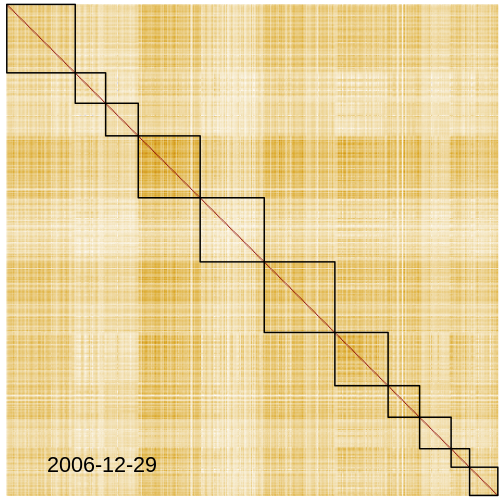}
  \end{minipage}
  \begin{minipage}[b]{0.32\textwidth}
    \includegraphics[width=\textwidth]{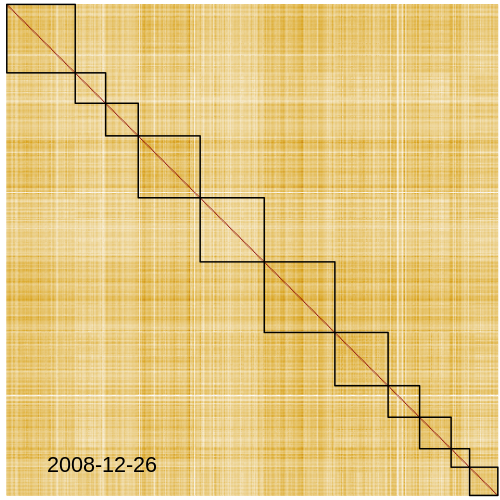}
  \end{minipage}
   \begin{minipage}[b]{0.32\textwidth}
    \includegraphics[width=\textwidth]{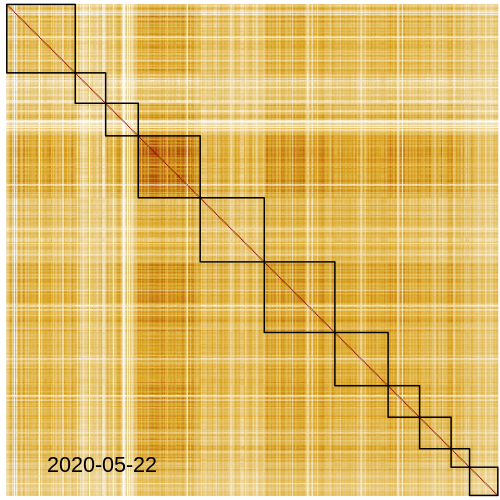}
  \end{minipage}
   \caption{\textit{Time-varying predicted mean return Correlations. Colors scale follows the same pattern from Figure \ref{factors_cov}.}}
   \label{ret_corrs}
\end{figure}

The evolution of correlations among stock returns is illustrated in Figure \ref{ret_corrs}, where we display the predicted mean correlation matrix for the selected periods. Stock returns are grouped according and following the same ordering as in Table \ref{sectors}. Considering the end of 2006, the sparser factor loadings in Figure \ref{fig: factor_loadings} and weaker factor correlations of Figure \ref{factors_cov} are translated into a lighter correlation matrix of returns. Again, as lighter the colours the closer to zero the correlations are. Note a sligthly stronger correlation among stocks from the Financial sector and how they are also more correlated to almost all other sectors. This patterns are repeated for the three selected periods. At the other hand, we can note how stocks from the Health Care sector are much less correlated with other stock returns. When we focus on the stressed periods of the Global Financial Crisis and the Covid-19 pandemic, the higher factor loadings and greater factor correlations are translated in higher stock return correlations. It is an expected pattern, since in bad market periods stock returns variations tend to be more related to factor movements. Interestingly, during the Covid-19 pandemic the Consumer Staples sector and some companies from the Energy sector  presented a strong correlation reduction, while the Financial sector is very high correlated not just among its own companies but with other sectors.

\begin{figure}[!h]
\begin{center}

	\includegraphics[width=10cm]{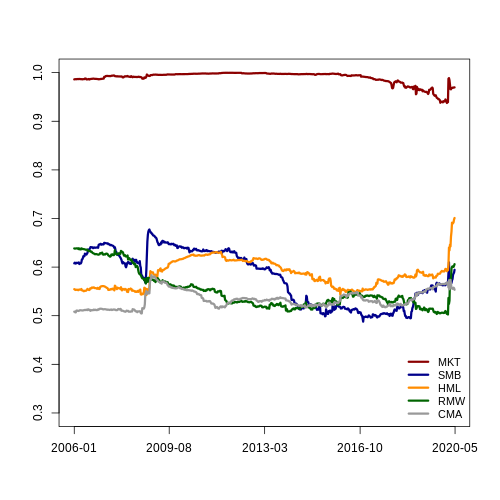} 
\caption{\textit{Time-varying inclusion probabilities for the five Fama-French risk factors over time generated from the Dynamic Risk Factor Dependency Model. Factor inclusion probabilities are represented as the cross-sectional average of inclusion probabilities for all stock returns.}}
\label{fig: pip}
\end{center}
\end{figure}

Finally, Figure \ref{fig: pip} investigates the evolution of risk factors inclusion probabilities. It can be defined as the posterior probability of an asset return including a specific covariate variable. Hence, for a given asset return $j$, the inclusion probability of a risk factor can be computed as the sum of probabilities of all univariate models for return $j$ including this specific risk factor. Since each asset return has different inclusion probabilities for different risk factors, we display in Figure \ref{fig: pip} the cross-sectional average of inclusion probabilities for each period of time. We can notice that the probability of including the market factor is considerably higher than all other risk factors, which demonstrates its importance in dictating common movements in the cross-section of stock returns. Since the market factor inclusion probability always fluctuates slightly close to 100\%, we conclude that models including this factor tend produce much stronger predictive densities compared to models excluding the traditional market factor. It is interesting to note a slight drop in the importance of the market factor since 2016, but with the advent of the Covid-19 pandemic, its importance returned to past values. Indeed, except the CMA factor, all other risk factors have become substantially more relevant since the beginning of the pandemic. This increase in importance was also observed during the Great Recession for SMB, HML and CMA. The general assessment after the Great Recession is that the HML factor demonstrated a higher inclusion probability than the SMB, RMW and CMA factors for the great majority of the time. Therefore, Figure \ref{fig: pip} clearly shows how our DRFDM is able to infer about the importance of different risk factors in a dynamic fashion, providing evidences of a time-varying impact of different risk factors on stock returns.

In what follows in the next subsections, we study analysis of several variants and restrictions of our DRFDM based on different discount factor specifications and risk factors to be considered. We let the first four years of data (from 2002 to 2005) as a training period and we perform statistical and portfolio out-of-sample evaluation for the next years. Therefore, we discard the first 208 data points to train our models and use the next 751 data points for evaluation.

\paragraph{Model specifications:}Before going to the statistical and portfolio analysis, we detail here several model variants and restrictions to be compared. Different model settings are based on previous tests and experience using the initial training sample and values are similar to the commonly applied by the econometric literature (\citealp{gruber2017bayesian} and \citealp{zhao2016dynamic}):  

\begin{itemize}
    \item DRFDM:  This model considers the 5 Fama-French factors (\citealp{fama2015five}) as asset returns \textit{parents}, applying dynamic risk factor and discount factors selection. Hence, it learns automatically the variation in betas (factor loadings), the variation in factor covariances, the variation in return volatilities and the selection of the best risk factors for each individual asset return and for each period of time. In terms of discount factors for different degrees of variation in coefficients, we let the model choose between: $\kappa_{r} \in \{0.99, 0.995, 1 \}$ for return volatilities; $\delta \in \{0.998, 0.999, 1 \}$ for factor loadings; and $\kappa_{f} \in \{0.999, 1 \}$ for factor volatilities. Model probabilities are computed using a forgetting factor $\alpha=0.99$.
    
    \item DRFDM ($\alpha=0.98$): The same as DRFDM, but considering $\alpha=0.98$ for model probabilities.
    
    \item DRFDM ($\alpha=1$): The same as DRFDM, but considering $\alpha=1$ for model probabilities. In this case, it applies BMS, the static version of DMS.
    
    \item DRFDM (No Spars.): A dense version of DRFDM. It does not induce sparsity by dynamic factor selection. Therefore, all 5 Fama-French factors are always considered. 
    
    \item 3F-DRFDM: The same as DRFDM, but restricting the set of risk factors to MKT, SMB and HML (\citealp{fama1992cross}).
    
    \item 4F-DRFDM: The same as 3F-DRFDM, but including MOM as an additional risk factor (\citealp{carhart1997persistence}).
    
    \item 6F-DRFDM: The same as DRFDM, but including MOM as an additional risk factor.

    \item W-DLM: It is the standard multivariate Wishart-DLM (see \citealp{prado2010time}, Ch. 10). We use $\delta=0.997$ for the local level evolution discounting and $\kappa = 0.99$ for the multivariate volatility discount factor.\footnote{For both W-DLM and Factor W-DLM we set the initial prior location $S_{0}$ to a diagonal matrix whose diagonal elements are given by 0.1. We also tested using the residual variances over the training period from an OLS model, but using 0.1 instead produced a stronger benchmark. \cite*{moura2020comparing} have shown that using a Wishart process with a shrinkage towards a diagonal covariance matrix delivers better portfolio performance with lower turnovers.} 
    
    \item Factor W-DLM: It is a multivariate Wishart-DLM using the 5 Fama-French factors as observable common predictors. We have considered $\delta=0.997$ for states discounting and $\kappa = 0.99$ for the multivariate volatility discount factor.\footnote{We model risk factors in a separate W-DLM and predict asset returns and covariances conditional on risk factor predictions.}
        
\end{itemize}

 We also show different combination of models still applying dynamic factor selection, but considering the following restrictions:

 \begin{itemize}
    \item TVB: A model using time-varying betas (factor loadings), where $\delta = 0.999$.
    
    \item CB: A model using constant betas (factor loadings), where $\delta = 1$.
    
    \item FSV: A model using factor stochastic volatility, where $\kappa_{f} = 0.999$.
    
    \item FCV: A model using factor constant volatility, where $\kappa_{f} = 1$.
    
    \item SV: A model using return stochastic volatility, where $\kappa_{r} = 0.995$.
    
    \item CV: A model using return constant volatility, where $\kappa_{r} = 1$.

\end{itemize}

As we explain in Appendix A, discount factors equal to one represent the case of no variation in coefficients, while discount factors lower than one induce time-variability in coefficients. Hence, the DRFDM is the most flexible model, since it allows to switch between constant and time-varying parameters and dynamically selects different subset of risk factors over time if it is empirically desirable.

\subsection{Forecasting accuracy }
\label{statistical}

In order to evaluate different model settings in terms of statistical accuracy, we compute density forecasts and hit rates. As we explain in details in Appendix B, for each period of time an individual asset return $j$ can load in different subsets of risk factors \textit{parents}. Hence, in order to generate an asset return forecasting for the next week, the investor considers the predictions obtained from the specific risk factor parents for that asset and prior coefficients:

$$
f_{jt|t-1} = a_{j \alpha t}+\lambda_{pa(j) t|t-1}^{\prime} a_{j \beta t}
$$

\noindent where $f_{jt|t-1}$ is the asset return $j$ forecasting for the next week and $\lambda_{pa(j) t|t-1}$ represents a vector of risk factor predictions. It is important to highlight here that for our \textit{Dynamic Risk Factor Dependency Model}, this set of risk factor dependencies changes depending on the period of time and for different asset returns under analysis. Finally, $a_{j \alpha t}$ and $a_{j \beta t}$ are the prior means for time $t$ for, respectively, the intercept and factor loading distributions given all data available until time $t-1$.

In Table \ref{stat_performance} we show sign accuracy (Acc.) and the Log-Predictive Density (LPD)\footnote{We exclude from the analysis the measures related to risk factors, focusing only on Acc. and LPD for the $N$ asset returns, which is the main interest of our portfolio allocation exercise.}. The first is computed as the sum of the log of the 1-step ahead predictive densities over the evaluation period as in Equation (\ref{eq:predictive}) for the $N$ assets available, and higher values represent better performance. The LPD gives a sense of how well a model performs out-of-sample considering its whole predictive distribution and not just its mean. Therefore, it suits quite well to our portfolio analysis, since it also takes into account the impact of the covariance structure. This metric has been applied recently in many bayesian econometric papers and has become common practice for model comparison (see \citealp{koop2013large}, \citealp*{zhou2014bayesian}, \citealp{gruber2017bayesian} and \citealp*{mcalinn2020multivariate}). The second statistical metric is simply computed as the number of corrected sign predictions averaged across all assets and over the out-of-sample evaluation period. Hence, we represent the the accuracy of model $k$ over the evaluation period as

$$
Acc_{k} =  \frac{\sum_{j} \sum_{t=(T_{0}+1)}^{T}  \mathbbm{1}_{\{sign(f_{jt|t-1}^k) = sign(r_{jt})\}}}{(T-T_{0})N} 
$$

\noindent where $T_{0}$ represents the training period and $\mathbbm{1}_{\{sign(f_{jt|t-1}^k) = sign(r_{jt})\}}$ is an indicator function, which is equal to one when the sign of forecasted asset return $j$ for period $t$ using model $k$ is the same sign of the actual observed asset return $j$ for period $t$.\footnote{A sign is equal to 1 when returns are positive and equal to -1 when are negative.}

Table \ref{stat_performance} provides statistical out-of-sample performances. First, what can be noticed is that all different settings are performing much better than the Wishart approach in terms of density forecast. In general, models with constant parameters produce worst density forecast as well, specially with CV.  In terms of the number of factors to be considered, the density forecast measure is quite similar among different choices, with a slight worsening for models including the CMA and RMW factors. In terms of correct return sign forecasts, differences are quite small. However, the recent literature on empirical asset pricing has shown evidence that even models providing only a small improvement in return predictability are capable to generating significant impacts on portfolio performance over time (\citealp*{chinco2019sparse}, \citealp*{gu2020empirical}, \citealp*{jiang2020re} and \citealp{levy2021time}). For instance, one may note in Table \ref{stat_performance} that the model with no sparsity on factor loadings produced very similar statistical performance than those models inducing sparsity, with just a small statistical deterioration. As we show in the empirical portfolio analysis (section \ref{port_performance}), achieving sparsity promotes strong mean and variance prediction stability. In a high-dimension portfolio allocation with thousands of factor loadings for each period of time, a reduction on the parameter space and small differences on predictions are able to generate great improvements on final portfolio performance. We can also conclude from Table \ref{stat_performance} that models with constant factor loadings tend to produce worse out-of-sample accuracy in terms of both predictive densities and hit rates.

\begin{table}[!h] \centering 
  \caption{Out-of-sample forecasting accuracy} 
  \label{stat_performance} 
\begin{tabular}{@{\extracolsep{5pt}}l ccc} 
\\[-1.8ex]\hline 
\hline \\[-1.8ex] 
 & LPD & Acc. \\ 
\hline \\[-1.8ex] 
DRFDM & $894.7$ & $53.21$ \\ 

DRFDM ($\alpha=1$) & $894.2$ & $53.18$ \\ 
DRFDM ($\alpha=0.98$) & $893.2$ & $53.19$ \\ 
DRFDM (No Spars.) & $894.4$ & $53.18$ \\ 
3F-DRFDM & $896.5$ & $53.20$ \\ 
4F-DRFDM& $895.2$ & $53.15$ \\ 
6F-DRFDM & $893.4$ & $53.19$ \\ 

TVB-FSV-CV & $856.0$ & $53.17$ \\ 
TVB-FCV-CV & $856.0$ & $53.17$ \\ 
TVB-FSV-SV & $879.5$ & $53.17$ \\ 
TVB-FCV-SV & $879.5$ & $53.17$ \\ 
CB-FSV-SV & $878.0$ & $53.16$ \\ 
CB-FSV-CV & $854.2$ & $53.16$ \\ 
CB-FCV-CV & $854.2$ & $53.14$ \\ 
CB-FCV-SV & $878.0$ & $53.14$ \\ 
Factor-W-DLM & $777.1$ & $53.21$ \\ 
W-DLM & $780.5$ & $53.19$ \\ 
Mom Signal &- & 52.25\\
\hline \\[-1.8ex] 
\end{tabular} 
\begin{tablenotes}
      \small 
    \item    \textit{The Table reports out-of-sample forecasting performance for different model settings. The first column displays Log-Predictive Densities (LPD) as a measure of density forecasting. The second column shows hit rates (Acc.), represented as the average number of correctly forecasted return signs.  
}
    \end{tablenotes}
\end{table}

It is important to notice that even though both W-DLM and Factor W-DLM provide much weaker predictive densities than other model settings, they show stronger out-of-sample accuracy in terms of hit rates, with the Factor W-DLM showing the same accuracy as the main DRFDM. However, as we show in the empirical portfolio allocation section, both Wishart approaches produce portfolios with performances considerably lower than our DRFDM. In fact, this results confirm the evidence obtained in \cite{cenesizoglu2012return}, where the authors show stronger correlation between density forecast and final portfolio performance than with point forecast metrics. Since any point forecast metric is not able to incorporate any uncertainty around predictions and the covariance structure among returns, it tends to generate a lower correlation with final portfolio performances. 

Finally, at the bottom of Table \ref{stat_performance} we provide the hit rate using the momentum signal as a return predictor. In this case, instead of using an econometric model, the investor only considers the momentum from previous 12 months as a measure to predict future returns. This type of approach has received increasing attention in the recent literature and has been applied to high-dimensional portfolio problems (\citealp*{engle2019large}, \citealp*{de2018factor} and \citealp{moura2020comparing}).\footnote{Since we use weekly data, we consider the momentum as the average of previous 52 returns on each stock but excluding the most recent 4 returns.}


\subsection{Dynamic Portfolio Allocation}

After describing the set of possible models considered in this study and their statistical performance, we discuss how our DRFDM is able to improve final investor decisions. We will take the perspective of an investor who allocates her wealth among all different stocks available in the dataset. At each period of time, the investor applies two steps. The first is to use the econometric method to generate one-week ahead forecasts of return means and covariances. In the second step, she dynamically rebalances the portfolio by finding new optimal portfolio weights. In our main analysis, we perform a Mean-Variance portfolio optimization, where the investor uses both the vector of predicted mean stock returns and its respective predicted covariance matrix, as we describe below. Although it is not the focus of our analysis, in Appendix C we also show additional results for the case where the investor only considers the predicted covariance matrix for a Global Minimum Variance portfolio. Hence, from this setup we are able to assess the economic value of the DRFDM with different settings within a dynamic framework, implementing an efficient-frontier strategy subject to return target or a minimum-risk portfolio. Below we describe in more details the portfolio strategies.

\subsubsection{Mean-Variance Portfolio}

The Mean-Variance Portfolio (MVP), also know as Efficient Frontier (EFF) portfolio or Markowitz portolio due to the seminal work of \cite{markowitz1952portfolio}, solves the following investment problem in the absence of short-sales constraints:

\begin{equation}
\begin{gathered}
\label{mvp_opt}
\min _{\boldsymbol{\omega}_{t}} \boldsymbol{\omega}_{t}^{\prime} \boldsymbol{\Sigma_{t}} \boldsymbol{\omega}_{t} \\
\text { subject to } \quad \boldsymbol{\omega}_{t}^{\prime} \boldsymbol{\mu}=\tau, \text { and } \\
\boldsymbol{\omega}_{t}^{\prime} \mathbf{1}=1
\end{gathered}
\end{equation}

\noindent where $ \mathbf{1}$ is a vector of ones, $\boldsymbol{\Sigma_{t}}$ is the covariance matrix of returns for time $t$, $\boldsymbol{\omega}_{t}$ represents portfolio weights, $\boldsymbol{\mu}$ is the expected return vector and $\tau$ is the return target. Replacing $\boldsymbol{\mu}$ by the predicted vector of returns from our econometric approach $\boldsymbol{f}_{t|t-1} = \boldsymbol{\alpha}_{t|t-1} + \boldsymbol{\beta}_{t|t-1} \boldsymbol{\lambda}_{t|t-1}$\footnote{More details about predictive moments computation can be found in Appendix B.}, and $\boldsymbol{\Sigma_{t}}$ by the respective point estimate of the predicted covariance matrix, $\widehat{\boldsymbol{\Sigma}}_{t|t-1}$, the expression for the optimal portfolio weights can be expressed as:

\begin{equation}
\label{w_eff}
\boldsymbol{\omega}_{t} =\frac{C-\tau B}{A C-B^{2}} \widehat{\boldsymbol{\Sigma}}_{t|t-1}^{-1}  \mathbf{1}+\frac{\tau A-B}{A C-B^{2}} \widehat{\boldsymbol{\Sigma}}_{t|t-1}^{-1} \boldsymbol{f}_{t|t-1}
\end{equation}

\vspace{3mm}

\noindent where $A=  \mathbf{1}^{\prime} \widehat{\boldsymbol{\Sigma}}_{t|t-1}^{-1}  \mathbf{1}$, $B= \mathbf{1}^{\prime} \widehat{\boldsymbol{\Sigma}}_{t|t-1}^{-1}\boldsymbol{f}_{t|t-1}$ and $C=\boldsymbol{f}_{t|t-1}^{\prime} \widehat{\boldsymbol{\Sigma}}_{t|t-1}^{-1} \boldsymbol{f}_{t|t-1}$ . Following \cite{engle2012dynamic}, in our main results we have considered  an annualized return target of $\tau = 10\%$, but we also report additional results for $\tau = 15\%$ and $20\%$.

In Appendix C we also show results for a restricted portfolio optimization, where we solve a problem similar to (\ref{mvp_opt}), but including one additional restriction: the maximum (absolute) weights on individual stocks to be 5\%.

\subsubsection{Global Minimum Variance}

In Appendix C we show as additional analysis the results for the Global Minimum Variance portfolio (GMV) with no short-sales constraints. The main investment problem of the GMV is to reduce total portfolio risk and can be expressed as:

\begin{equation}
\min _{\boldsymbol{\omega}_{t}} \boldsymbol{\omega}_{t}^{\prime} \boldsymbol{\Sigma_{t}} \boldsymbol{\omega}_{t} \quad \text { s.t. } \boldsymbol{\omega}_{t}^{\prime} \mathbf{1}=1 .
\label{min_var}
\end{equation}

\noindent where again $\boldsymbol{\Sigma_{t}}$ is the covariance matrix of returns for time $t$ and $\boldsymbol{\omega}_{t}$ represents a vector of portfolio weights. Replacing $\boldsymbol{\Sigma_{t}}$ by the predicted covariance matrix, the optimal portfolio weights for the GMP is given by:

 \begin{equation}
 \label{w_gmvp}
\boldsymbol{\omega}_{t}=\frac{\widehat{\boldsymbol{\Sigma}}_{t|t-1}^{-1}\mathbf{1}}{\mathbf{1}^{\prime} \widehat{\boldsymbol{\Sigma}}_{t|t-1}^{-1} \mathbf{1}}
\end{equation}
 
Since the GMV focus is to reduce portfolio volatility, it ignores the use of the mean of returns and tend to generate lower Sharpe ratios in practice. Although it is not widely used in practice, this approach can be viewed as a way to evaluate covariance matrix estimation and is still a common practice in academic works. Besides the GMV portfolio as in Equation (\ref{min_var}), we also show results when we include an additional restriction of maximum (absolute) weights on individual stocks to be 5\%, as we did on the MVP case.

\subsubsection{Economic Performance Measures}

After producing forecast outputs to dynamically build portfolios, we backtest our models in terms of portfolio performance out-of-sample. Investors face portfolio allocation problems and, at the end of the day, it is not just about out-of-sample predictability, but how predictions are translated into better final decisions, i.e., better portfolio choices. With this in mind, we evaluate portfolios based on financial metrics, such as portfolio turnovers, annualized mean excess returns (Mean), standard deviations (SD) and Sharpe ratios (SR). The latter is commonly used among practicioniers in the financial market and by academics. Despite their popularity, those portfolio metrics are unconditional measures and are not well suited for dynamic allocations with time-varying and sequential predictions (see \citealp{marquering2004economic}). Also, they do not take into account the investor risk aversion. In order to overcome this problems and improve our model comparisons we follow \cite*{fleming2001economic} and provide a measure of economic utility for investors.

We compute ex-post average utility for a mean-variance investor with a quadratic utility. As in \cite{fleming2001economic} and \cite*{della2009economic} we can calculate the performance fee that an investor will be willing to pay to switch from the traditional Wishart Dynamic Linear Model (W-DLM) benchmark to our \textit{Dynamic Risk Factor Dependency Model}. The performance fee is computed by equating the average utility of the W-DLM portfolio with the average utility of the DRFDM portfolio (or any other alternative portfolio), considering the latter with a management fee $\Phi$:

\begin{equation}
  \small \small \sum_{t=0}^{T-1}\{(R_{p, t+1}^{DRFDM}-\Phi)-\frac{\gamma}{2(1+\gamma)}(R_{p, t+1}^{DRFDM}-\Phi)^{2}\} =  \sum_{t=0}^{T-1}\{R_{p, t+1}^{W-DLM}-\frac{\gamma}{2(1+\gamma)}(R_{p, t+1}^{W-DLM})^{2} \}  
\end{equation}

\noindent where $\gamma$ is the investor's degree of relative risk aversion and $R_{p,t}^{DRFDM}$ is the gross excess return of the DRFDM portfolio and $R_{p,t}^{W-DLM}$ is the gross excess return from the W-DLM portfolio. As in \cite{fleming2001economic}, we report our estimates of $\Phi$ as annualized fees in basis points using $\gamma = 10$.\footnote{Appendix C provides additional results applying different values for $\gamma$ and main conclusions are maintained.} 

All economic measures displayed in Section \ref{port_performance} are already net of transaction costs (TC). Following \cite{marquering2004economic}, we deduct  transaction costs from the portfolio return ex-post. Despite the great majority of the papers related to covariance matrix estimation for portfolio allocation cited above do not take into account transaction costs in their findings, in our main results we consider TC = 5 bps of the traded volume in an effort to approximate our results to a real world example. In general, there is disagreement about which transaction cost to incorporate. In the past, many papers applied transaction costs of 50 bps, but recently this value has been substantially reduced for the most liquid stocks (\citealp{french2008presidential}). Hence, in the same spirit of \cite*{moura2020comparing} we display additional results for TC $\in \{0, 10 \}$ bps.

\subsubsection{Additional Competing Models}

Here we detail some additional covariance matrix estimation approaches included in our portfolio performance analysis. We consider some recent traditional benchmarks from the literature. We also show results for two benchmarks that do not require the use of econometric models to be estimated: the equally-weighted portfolio from our universe of stocks and the passive investment in the $S\&P$ 500 index. The latter can be viewed as a strong benchmark, since it is well known that is quite hard to beat the market.

\begin{itemize}
    \item EFM: this is a static estimator based on an exact factor model, where $\boldsymbol{\Sigma}_{f}$ is given by the sample covariance matrix of risk factors and the residual covariance matrix $\boldsymbol{\Omega}$ is a diagonal matrix filled with the sample variances estimates of residuals. We estimate residuals and factor loadings using four years rolling regressions. 
    
    \item DCC-NL: it is a dynamic estimator using the multivariate GARCH of \cite{engle2019large}.
    
    \item AFM-DCC-NL: it is the dynamic approximate factor model of \cite*{de2018factor}. \footnote{Following the main results from their paper, we have consider only the market factor.} 
    
    \item LW: it is the static linear shrinkage estimator of \cite{ledoit2004honey}.
    
    \item EWMA: the traditional exponentially weighted moving average estimator where $\boldsymbol{\Sigma}_{t+1}=(1-\lambda) \boldsymbol{y}_{t}^{\prime} \boldsymbol{y}_{t}+\lambda \boldsymbol{\Sigma}_{t}$. We consider two values for the decay factor, $\lambda \in \{0.97, 0.99\}$.

    \item EW:  a simple strategy considering a equal-weighted portfolio over our universe of stocks. As claimed by \cite*{demiguel2009optimal}, it tends to perform better than the simple unconditional covariance matrix of returns and it has been claimed to be hard to be outperformed. 
    
    \item $S\&P$: it represents the passive investment (buy-and-hold) on the $S\&P$ 500 index over the evaluation period.

\end{itemize}

It is important to highlight that for models EFM, DCC-NL, AFM-DCC-NL, LW and EWMA detailed above, we use the average momentum signal from the previous 12 months as vector of predicted mean returns to find optimal portfolio weights in Equation (\ref{w_eff}). It follows a similar procedure as applied in \cite{engle2019large}, \cite{de2018factor} and \cite{moura2020comparing} and can be viewed as a competitive method to forecast returns.

We recognize the existence of other recent competing models in the literature involving Bayesian analysis using MCMC methods, as we have described in Section \ref{sec1: Introduction}. However, the simulation schemes dramatically limit scalability and require repeat MCMC simulation analyses at each time point which makes the whole backtesting procedure computationally prohibitive. Although factor stochastic volatility models of \cite{kastner2017efficient} and \cite{kastner2019sparse} have been implemented in the R package \textit{factorstochvol}, \footnote{See \cite{hosszejni2019modeling}} the estimation is not sequential, which requires the model to be rerun at each time over the evaluation period. As argued by \cite{gruber2017bayesian}, it will take several weeks to backtest a universe of stocks like ours. At the other hand, since our DRFDM estimation is sequential and does not require any simulation schemes, it is able to handle the whole estimation procedure in only few minutes.

\subsection{Portfolio Performance}
\label{port_performance}

In order to evaluate the validity of the DRFDM out-of-sample in a real world context, we conduct a backtest analysis using the different portfolio strategies and model settings described in the previous sections. One of the great advantages of our approach is its flexibility and fast computation. For the sake of curiosity,
using parallel computations among 32 cores we run our main model (DRFDM with 5 Fama-French factors) for the entire dataset and produced portfolio backtests in less than 10 minutes.\footnote{We also have repeated the exercise using almost one thousand stocks from the Russel 1000 Index and we were able to run the whole estimation procedure and portfolio backtest in less than 25 minutes. }

\begin{table}[!h] \centering 
  \caption{Mean-Variance Portfolio performances (TC= 5 bps) } 
  \label{eff_tc5} 
  \small
\begin{tabular}{@{\extracolsep{5pt}}l cccccc} 
\\[-1.8ex]\hline 
\hline \\[-1.8ex] 
 & Turnover & Mean & SD & SR & $\Phi$ \\ 
\hline \\[-1.8ex] 
EFM & $0.23$ & $6.9$ & $13.3$ & $0.52$ & $136.2$ \\ 

AFM-DCC-NL & $0.53$ & $6.6$ & $11.3$ & $0.58$ & $355.5$ \\ 

DCC-NL & $0.97$ & $6.1$ & $12.3$ & $0.50$ & $189.8$ \\ 
LW & $0.23$ & $7.1$ & $11.9$ & $0.59$ & $333.7$ \\ 
EWMA(0.99) & $0.80$ & $4.7$ & $14.2$ & $0.33$ & $-202.7$ \\ 
EWMA(0.97) & $2.07$ & $7.2$ & $20$ & $0.36$ & $-929.9$ \\ 
DRFDM & $0.19$ & $10.5$ & $12.8$ & $0.82$ & $585.1$ \\ DRFDM (No Spars.)& $0.14$ & $10.7$ & $13.7$ & $0.78$ & $467.7$ \\ 
DRFDM ($\alpha=1$)& $0.14$ & $10.7$ & $13.5$ & $0.79$ & $495.8$ \\ 
DRFDM ($\alpha=0.98$)& $0.23$ & $9.0$ & $12.4$ & $0.73$ & $477$ \\ 
3F-DRFDM & $0.15$ & $10.4$ & $13.1$ & $0.80$ & $530.1$ \\ 
4F-DRFDM & $0.18$ & $10.6$ & $13.1$ & $0.81$ & $541.7$ \\ 
6F-DRFDM & $0.21$ & $10.4$ & $12.8$ & $0.81$ & $568.4$ \\ 
DRFDM (Mom signal) & $0.24$ & $10.0$ & $12.5$ & $0.80$ & $572.3$ \\ 
TVB-FSV-CV & $0.15$ & $8.9$ & $13.4$ & $0.66$ & $329.8$ \\ 
TVB-FCV-CV & $0.16$ & $8.9$ & $13.4$ & $0.66$ & $331.6$ \\ 
TVB-FSV-SV & $0.17$ & $9.6$ & $13$ & $0.74$ & $453.6$ \\ 
TVB-FCV-SV & $0.18$ & $9.6$ & $13$ & $0.74$ & $456.5$ \\ 
CB-FSV-SV & $0.17$ & $9.6$ & $13$ & $0.74$ & $456.4$ \\ 
CB-FSV-CV & $0.15$ & $8.8$ & $13.4$ & $0.65$ & $311.7$ \\ 
CB-FCV-CV & $0.15$ & $8.8$ & $13.4$ & $0.65$ & $313$ \\ 
CB-FCV-SV & $0.17$ & $9.6$ & $13$ & $0.74$ & $459.2$ \\ 

Factor-W-DLM & $0.69$ & $5.4$ & $13.6$ & $0.40$ & $-56.1$ \\ 
W-DLM & $0.66$ & $5.7$ & $13.4$ & $0.43$ & $0.0$ \\ 
EW & $0.03$ & $7.1$ & $21.8$ & $0.32$ & $-1,180.2$ \\ 
S\&P &  & $8.3$ & $18.2$ & $0.46$ & $-507$ \\ 
\hline \\[-1.8ex] 
\end{tabular} 
\begin{tablenotes}
      \small 
    \item    \textit{The Table reports out-of-sample portfolio performances for several covariance models following optimal portfolio weights from Equation (\ref{w_eff}) and using an annualized return target of $\tau=10\%$. Mean excess returns (Mean), volatilities (SD), Sharpe Ratios (SR) and management fees ($\Phi$) are reported in annual terms whereas turnover are in weekly terms. Results are computed using returns net of transaction costs of 5 bps. Annualized management fees are computed considering a relative risk aversion of $\gamma=10$. 
}
    \end{tablenotes}
\end{table}

The main focus of our portfolio analysis is to find an econometric model able to satisfactorily handle the best balance between risk and return. Table \ref{eff_tc5} present results related to the MVP portfolio with 5 bps transaction costs during the out-of-sample evaluation period (from January, 6, 2006 until May, 22, 2020). The main columns to be analyzed from this table are in terms of SR and annualized management fee ($\Phi$). What can be noticed from Table \ref{eff_tc5} is that regardless of the selected model setting, all models outperform the EW portfolio suggested by \cite{demiguel2009optimal}. It is interesting to notice that although the EW strategy presents the lowest weekly turnover, it has the worst SR because of its high volatility that is caused by the absence of a covariance structure among returns on its formulation. Also, the great majority of the models were able to outperform the passive investment on the $S\&P$500 index. In special, the SR from the DRFDM and its different subsets of risk factors counterparts dominate all other models, with DRFDM presenting an out-of-sample SR of 0.82, almost two times the SR from the $S\&P$500 index, the Factor W-DLM and the W-DLM benchmark. The latter performed quite similar to the $S\&P$500 index in terms of SR, but since it was able to produce much lower portfolio volatility, it generates utility gains for the investor compared to the market index. However, since our DRFDM was able to reduce volatilty even further and produce quite stable return predictions, it has produced a considerable utility gain for the investor. In fact, a mean-variance investor would be willing to pay an annualized management fee of 585 bps to switch fromm the traditional W-DLM benchmark to the DRFDM approach. Using different subsets of risk factors in our DRFDM setting also produced quite strong portfolio results.

We also included in our analysis what we call DRFDM (Mom signal), which represents the same model as the main DRFDM, but instead of using its predicted mean returns for portfolio optimization in Equation (\ref{w_eff}), we used the same momentum signal as we did for competing models at the top of Table \ref{eff_tc5}. What we observe is that using the predictions from our approach generates more stable estimates and portfolio performance than using the momentum signal. Although the differences are small, the DRFDM (Mom signal) induces higher turnovers, what harms final portfolio performances as higher transaction costs start to be considered, as Table \ref{eff_tc10} shows. Also, as Table \ref{stat_performance} had demonstrated, predictions from the DRFDM generated higher out-of-sample accuracy than the Momentum Signal. Hence, we see the mean predictions from our approach as a clear competitor to the classical Momentum Signal broadly used in the academic literature and among practitioners.

In terms of the benefits of inducing sparsity on the covariance structure of returns, we see from Table \ref{eff_tc5} that the DRFDM (No Spars.) produced a much more volatile portfolio than when we allow to dynamically select the best risk factors for each stock return, which is translated in lower utility gains and SR for the investor. The benefits of time-varying sparsity on factor loadings are observed in all different portfolio setting of this paper, regardless of the optimization problem, amount of transaction costs or risk aversions. When the DRFDM set many factor loadings to exactly zero, it is indeed improving covariance matrix estimation by deflating its whole structure.

\begin{table}[!h] \centering 
  \caption{Mean-Variance Portfolio performances (TC= 10 bps)} 
  \small
  \label{eff_tc10} 
\begin{tabular}{@{\extracolsep{5pt}}l cccccc} 
\\[-1.8ex]\hline 
\hline \\[-1.8ex] 
 & Turnover & Mean & SD & SR & $\Phi$ \\ 
\hline \\[-1.8ex] 
EFM & $0.23$ & $6.3$ & $13.3$ & $0.47$ & $250.3$ \\ 

AFM-DCC-NL & $0.53$ & $5.2$ & $11.3$ & $0.46$ & $389.8$ \\ 

DCC-NL & $0.97$ & $3.6$ & $12.3$ & $0.29$ & $107.4$ \\ 
LW & $0.23$ & $6.5$ & $11.9$ & $0.54$ & $448.5$ \\ 
EWMA(0.99) & $0.80$ & $2.6$ & $14.1$ & $0.18$ & $-239.2$ \\ 
EWMA(0.97) & $2.07$ & $1.8$ & $20.0$ & $0.09$ & $-1,257$ \\ 
DRFDM & $0.19$ & $10.0$ & $12.8$ & $0.78$ & $712.8$ \\
DRFDM (No Spars.)& $0.14$ & $10.3$ & $13.7$ & $0.75$ & $609.9$ \\ 
DRFDM ($\alpha=1$)& $0.14$ & $10.3$ & $13.5$ & $0.76$ & $636.9$ \\ 
DRFDM ($\alpha=0.98$)& $0.230$ & $8.4$ & $12.4$ & $0.68$ & $592.3$ \\ 
3F-DRFDM & $0.15$ & $10.0$ & $13.1$ & $0.77$ & $668.6$ \\ 
4F-DRFDM & $0.18$ & $10.1$ & $13.1$ & $0.77$ & $674$ \\ 
6F-DRFDM & $0.21$ & $9.9$ & $12.8$ & $0.77$ & $690.9$ \\ 
DRFDM (Mom signal) & $0.24$ & $9.4$ & $12.5$ & $0.750$ & $687.6$ \\

TVB-FSV-CV & $0.15$ & $8.5$ & $13.4$ & $0.63$ & $465.4$ \\ 
TVB-FCV-CV & $0.16$ & $8.5$ & $13.4$ & $0.63$ & $466.9$ \\ 
TVB-FSV-SV & $0.17$ & $9.1$ & $13.0$ & $0.70$ & $585.5$ \\ 
TVB-FCV-SV & $0.18$ & $9.1$ & $13.0$ & $0.70$ & $588.1$ \\ 
CB-FSV-SV & $0.17$ & $9.2$ & $13.0$ & $0.71$ & $590.7$ \\ 
CB-FSV-CV & $0.15$ & $8.4$ & $13.4$ & $0.62$ & $448.7$ \\ 
CB-FCV-CV & $0.15$ & $8.4$ & $13.4$ & $0.63$ & $449.8$ \\ 
CB-FCV-SV & $0.17$ & $9.2$ & $13.0$ & $0.71$ & $593.2$ \\ 

Factor-W-DLM & $0.690$ & $3.6$ & $13.6$ & $0.27$ & $-64$ \\ 
W-DLM & $0.66$ & $4.0$ & $13.4$ & $0.30$ & $0.0$ \\ 
EW & $0.03$ & $7.0$ & $21.8$ & $0.32$ & $-1,029.1$ \\ 
S\&P &  & $8.3$ & $18.2$ & $0.460$ & $-337.6$ \\ 
\hline \\[-1.8ex] 
\end{tabular} 
\begin{tablenotes}
      \small 
    \item    \textit{The Table reports out-of-sample portfolio performances for several covariance models following optimal portfolio weights from Equation (\ref{w_eff}) and using an annualized return target of $\tau=10\%$. Mean excess returns (Mean), volatilities (SD), Sharpe Ratios (SR) and management fees ($\Phi$) are reported in annual terms whereas turnover are in weekly terms. Results are computed using returns net of transaction costs of 10 bps. Annualized management fees are computed considering a relative risk aversion of $\gamma=10$. 
}
    \end{tablenotes}
    
\end{table}

We also compare different models within our model structure but restricting their variation in coefficients to follow the same pattern for all periods of time by fixing their discount factors. For instance, models with constant volatilities (CV) tend to produce portfolios with higher volatilities and lower SR and utility gains than models with time-varying volatilities. It is interesting to notice that models with both time-varying factor volatilities (FSV) and time-varying factor loadings (TVB) were not able to considerably improve portfolio performance, giving evidence of no significant predictive power increment by allowing time-varying betas and factor volatilities for all periods of time. At the other hand, since our DRDFM was allowed to dynamically select different discount factors over time, it was able to switch between periods of low and high variation on parameters, producing better adaptation to the data and stronger portfolio improvements. In terms of the selected forgetting factor $\alpha$ for model probabilities, we see an increase in portfolio volatility when no forgetting is applied ($\alpha=1$) - which is equivalent to a static Bayesian model selection approach - and comes with a lower SR and utility gain for the investor. However, when a higher forgetting is allowed using $\alpha=0.98$, a considerable portfolio volatility reduction is observed, but worse mean excess returns and higher turnovers are produced, affecting final performance. Hence, despite models with different $\alpha$'s  are still delivering robust results and  are able to outperform several competitor models, it seems that applying the intermediate value $\alpha=0.99$ generates a better balance between risk and return.

\begin{table}[!h] \centering 
  \caption{Mean-Variance Portfolio performances (No transaction costs)} 
  \small
  \label{eff_tc0} 
\begin{tabular}{@{\extracolsep{5pt}}l cccccc} 
\\[-1.8ex]\hline 
\hline \\[-1.8ex] 
 & Turnover & Mean & SD & SR & $\Phi$ \\ 
\hline \\[-1.8ex] 
EFM & $0.23$ & $7.5$ & $13.3$ & $0.56$ & $23.8$ \\ 

AFM-DCC-NL & $0.53$ & $8.0$ & $11.3$ & $0.70$ & $321.8$ \\ 
 
DCC-NL & $0.97$ & $8.7$ & $12.3$ & $0.70$ & $273.3$ \\ 
LW & $0.23$ & $7.7$ & $11.9$ & $0.64$ & $220.5$ \\ 
EWMA(0.99) & $0.80$ & $6.8$ & $14.2$ & $0.48$ & $-169.1$ \\ 
EWMA(0.97) & $2.07$ & $12.6$ & $20.0$ & $0.63$ & $-595.9$ \\ 
DRFDM & $0.19$ & $11.0$ & $12.8$ & $0.86$ & $459.2$ \\ 
DRFDM (No Spars.)& $0.14$ & $11.0$ & $13.7$ & $0.80$ & $327.7$ \\ 
DRFDM ($\alpha=1$)& $0.14$ & $11.1$ & $13.5$ & $0.82$ & $357$ \\ 
DRFDM ($\alpha=0.98$) & $0.23$ & $9.6$ & $12.4$ & $0.78$ & $363.4$ \\ 
3F-DRFDM & $0.15$ & $10.8$ & $13.1$ & $0.83$ & $393.7$ \\ 
4F-DRFDM & $0.18$ & $11.0$ & $13.1$ & $0.84$ & $411.5$ \\ 
6F-DRFDM & $0.20$ & $11.0$ & $12.8$ & $0.86$ & $447.7$ \\ 
DRFDM (Mom signal) & $0.24$ & $10.6$ & $12.5$ & $0.85$ & $458.7$ \\ 
TVB-FSV-CV & $0.15$ & $9.3$ & $13.4$ & $0.69$ & $196.4$ \\ 
TVB-FCV-CV & $0.16$ & $9.3$ & $13.4$ & $0.69$ & $198.4$ \\ 
TVB-FSV-SV & $0.17$ & $10.0$ & $13.0$ & $0.77$ & $323.7$ \\ 
TVB-FCV-SV & $0.18$ & $10.0$ & $13.0$ & $0.77$ & $326.8$ \\ 
CB-FSV-SV & $0.17$ & $10.0$ & $13.0$ & $0.77$ & $324.1$ \\ 
CB-FSV-CV & $0.15$ & $9.2$ & $13.4$ & $0.68$ & $176.9$ \\ 
CB-FCV-CV & $0.15$ & $9.2$ & $13.4$ & $0.68$ & $178.4$ \\ 
CB-FCV-SV & $0.17$ & $10.0$ & $13.0$ & $0.77$ & $327.2$ \\ 

Factor-W-DLM & $0.69$ & $7.2$ & $13.7$ & $0.53$ & $-48.2$ \\ 
W-DLM & $0.66$ & $7.5$ & $13.5$ & $0.55$ & $0.0$ \\ 
EW & $0.03$ & $7.2$ & $21.8$ & $0.33$ & $-1,328.4$ \\ 
S\&P &  & $8.3$ & $18.2$ & $0.46$ & $-670$ \\ 
\hline \\[-1.8ex] 
\end{tabular} 
\begin{tablenotes}
      \small 
    \item    \textit{The Table reports out-of-sample portfolio performances for several covariance models following optimal portfolio weights from Equation (\ref{w_eff}) and using an annualized return target of $\tau=10\%$. Mean excess returns (Mean), volatilities (SD), Sharpe Ratios (SR) and management fees ($\Phi$) are reported in annual terms whereas turnover are in weekly terms. Results are reported considering no transaction costs (TC = 0 bp). Annualized management fees are computed considering a relative risk aversion of $\gamma=10$. 
}
    \end{tablenotes}
\end{table}

When we focus on the competitor models at the top of Table \ref{eff_tc5}, we see lower SR and utility gains compared to our approach, jointly with much higher turnover. In special, the EWMA (0.97) produced large turnovers and volatility. Although the AFM-DCC-NL model presented lower SR and utility gains than DRDFM, it was the approach with lower annualized volatility, a pattern that is repeated throughout the different portfolio specifications in this paper. The DCC-NL and LW approaches were also able to deliver lower volatilities, but as the AFM-DCC-NL, they fail to produce a good balance between risk and return, delivering lower SR and utility gains. Since those approaches are quite unstable, they require the portfolio to be highly rebalanced over time, harming final performance. Figure (\ref{turnovers}) in the Appendix compares turnovers from our DRFDM to the dynamic estimators AFM-DCC-NL and DCC-NL and it is clear the ability of the DRDFM approach to produce more stable rebalances and reducing turnovers, specially during the Great Recessions and the Covid-19 pandemic. It also can be seen as an advantage of DRFDM for portfolio managers interested to invest in low liquid markets with much higher transaction costs.

Tables \ref{eff_tc10} and \ref{eff_tc0} repeat the same portfolio procedure as Table \ref{eff_tc5}, using different transaction costs. The conclusion are quite similar to those described before. However, when a higher transaction cost of 10 bps is considered, we can notice from Table \ref{eff_tc10} the considerable negative portfolio impacts on those competitor models with high turnovers, such as the DCC-NL, EWMAs, Factor W-DLM and W-DLM. Due to the characteristic of high diversification and low portfolio rebalancing changes from the DRFDM, it improves even further in terms of utility gains compared to the W-DLM benchmark, requiring the investor to pay an annualized management fee of 712 bps to change from the W-DLM to the DRDFM. When no transaction costs are considered, the W-DLM improves which reduce the utility gains of several models, including the DRFDM. However, our main model get a SR of 0.86, the same as the 6F-DRFDM and almost 23\% higher than the SR obtained from the AFM-DCC-NL and DCC-NL models.

In Appendix C the reader can find several additional tables reporting results using different return targets, risk aversions, portfolio constraints and applying the Minimum Variance Portfolio optimization of Equation (\ref{min_var}). In Table \ref{fees} we show annualized management fees when lower relative risk aversions are considered as a robustness analysis. In fact, the main conclusions remain the same for different levels of risk aversion, i.e., inducing time-varying sparsity and dynamically selecting different discount-factors for variation in coefficients by our DRFDM is able to considerably add  economic value for the investor compared to the traditional W-DLM benchmark and other common competitor models regardless of the risk aversion. Tables \ref{eff_15_tau} and \ref{eff_20_tau} show mean-variance portfolio performances using higher annualized portfolio return targets of 15\% and 20\%, respectively. What can be observed is that not only the conclusions are the same as those using a return target of 10\% from Table \ref{eff_tc5}, but are even stronger in terms of Sharpe Ratio and management fees the investor would pay to use the DRFDM. The same happens from the GMV optimization in Tables \ref{gmv_tc5}, \ref{gmv_tc10} and \ref{gmv_tc0} using different transaction costs. The DRFDM and its different specifications were able to deliver volatility reduction compared to the W-DLM and produce extremely lower turnovers and economic value for the investor. In spite of the fact that the DCC-NL, AFM-DCC-NL and LW have generated lower SD than our DRFDM, they fail to produce what is the advantage from our approach and is the main investor interest at the end of the day: better returns adjusted by risk and strong utility gains. Last but not least, Tables \ref{eff_w5} and \ref{gmv_w5} show mean-variance optimization and global minimum-variance portfolios with maximum weight constraints, where little impacts were observed to our approach because of the high diversification quality of the covariance matrix from the DRFDM, so the vast majority of portfolio weights were already below the 5\% limit in absolute terms when the formulas from Equations (\ref{mvp_opt}) and (\ref{min_var}) were being applied. Interestingly, competitor models with high turnovers and low portfolio diversification, such as EWMAs and Wishart models were able to considerably reduce portfolio turnovers and risks.

Summarizing, the main conclusions from different tables and results in this empirical section display similar informations: regardless of the parameters set by the investor, she is always benefited from the dynamic choices made by the DRFDM, producing lower risks, portfolio turnovers and utility improvements. Therefore, a mean-variance investor who dynamic learns about best risk factors, variation in coefficients and volatilities in an automatic and online fashion improves final decisions and utility measures compared to the benchmarks.

\section{Conclusion} \label{sec5: Conclusion}

Dynamic portfolio allocation in high-dimensions require a refined estimation of the covariance matrix of returns. The goal of this paper was to introduce a fast and flexible multivariate model for returns that is able to improve predictions for bayesian portfolio decisions. Inspired by the \textit{Cholesky-style} framework for multivariate inferences, we impose economically motivated risk factor dependencies that are able to solve the curse of dimensionality. Due to the low dimension of the risk factors, we are able to deal with the problem of ordering uncertainty in a sequential fashion. The conjugate format for foward filters and the use of discount factors for state evolutions and dynamic model probabilities allow our model to sequentially select the best specification choices for each asset return in parallel, such as best risk factors, degree of variation in factor loadings, volatilities and factor volatilities. We show that by the use of dynamic factor selection, we can achive higher model parsimony by time-varying sparsity on the parameter space in an online fashion. 

We have found that the \textit{Dynamic Risk Factor Dependency Model} is able to improve final portfolio decisions compared to the traditional W-DLM benchmark, the Equally Weighted portfolio, the passive investment on the $S\&P$ 500 index and many other competitor models in the literature. It generates not just risk reduction and better Sharpe ratios, but significant lower portfolio turnovers and utility gains for investors. We show that a mean-variance investor will be willing to pay a considerable management fee to switch from those strategies to the DRFDM approach. Also, since our approach does not require expensive MCMC schemes to draw from posterior distributions, we can backtest high-dimension portfolios in a matter of few minutes. This is good news for portfolio managers who are interested to improve investment strategies in a financial world with a large number of assets available, high model uncertainties and rapid and complex changes over time.

\clearpage

\appendix
\setcounter{secnumdepth}{0}

\section{Appendix A: Filtering and Forecasting}
\label{appendix A}

Following \cite{zhao2016dynamic} and \cite{levy2021dynamic}, we give details about the evolution and updating steps for the set of $K + N$ univariate DLMs.

\noindent \textbf{Posterior at $t-1$}:  At time $t-1$ and for each series $j$, we define the initial states for $\theta_{j t-1} $ and volatility $\sigma_{j t-1}$ as:

\begin{equation}
\theta_{j, t-1}, \sigma_{j t-1}^{-1} \mid \mathcal{D}_{t-1} \sim \mathcal{NG}\left(\boldsymbol{m}_{j, t-1}, \boldsymbol{C}_{j, t-1}, n_{j, t-1}, n_{j, t-1} s_{j, t-1}\right)
\label{joint}
\end{equation}

\vspace{5mm}

\noindent Equation (\ref{joint}) is the joint posterior distribution of model parameters at time $t-1$, known as a Normal-Gamma distribution. Hence, given the initial states, posteriors at $t-1$ evolve to priors at $t$ via the evolution equations:

$$
 \boldsymbol{\theta}_{j, t}=\boldsymbol{\theta}_{j, t-1}+\boldsymbol{\omega}_{j, t}, \quad \text { where } \quad \boldsymbol{\omega}_{j, t} \sim N\left(0, \mathbf{W}_{j, t} /\left(s_{j, t-1} \sigma_{j t}^{-1}\right)\right) 
 $$
 
 $$
 \sigma_{j t}^{-1}  = \sigma_{j t-1}^{-1}  \eta_{j, t} / \kappa_{j}, \quad \text { where } \quad \eta_{j, t} \sim \operatorname{Be}\left(\kappa_{j} n_{j, t-1} / 2,\left(1-\kappa_{j}\right) n_{j, t-1} / 2\right)
$$

\vspace{5mm}

\noindent where we can rewrite $\mathbf{W}_{j, t}$ as a discounted function of $\mathbf{C}_{j, t-1}$,   $\mathbf{W}_{j, t}=\mathbf{C}_{j, t-1}\left(1-\delta_{j}\right) / \delta_{j}$ for $0 < \delta_{j} \leq 1$ and the beta random variable $\eta_{j, t}$ is defined by the discount factor $0  < \kappa_{j} \leq 1$.  Discount methods are used to induce time-variations in the evolution of parameters and have been extensively used in many applications (\citealp{raftery2010online}, \citealp{dangl2012predictive}, \citealp{koop2013large}, \citealp{mcalinn2020multivariate} and others ) and well documented in \cite{prado2010time}. Note that lower values of $\delta$ and $\kappa$ induce higher degrees of variation in parameters and when discount factors are equal to one, both coefficients and volatilities will be constant.

Hence, the prior for time $t$ is given by 

\begin{equation}
\theta_{j t}, \sigma_{j t}^{-1} \mid \mathcal{D}_{t-1} \sim\mathcal{NG}\left(\boldsymbol{a_{j t}}, \boldsymbol{R_{j t}}, r_{j t}, r_{j t} s_{j, t-1}\right)
\end{equation}

\vspace{4mm}

\noindent where $r_{j t}=\kappa_{j} n_{j, t-1}$, $\boldsymbol{a_{j t}} = \boldsymbol{m_{j, t-1}}$ and  $\boldsymbol{R_{j t}} = \boldsymbol{C_{j, t-1}} / \delta_{j}$. 

\noindent \textbf{1-step ahead forecasts at time $t-1$}: The predictive distribution for $t$ at time $t-1$ given specif the risk factor \textit{parental set} for equation $j$ will be given by a Student's-$t$ distribution with $r_{j t}$ degrees of freedom:

$$
y_{j t} \mid y_{pa(j) t}, \mathcal{D}_{t-1} \sim \mathcal{T}_{r_{j t}}\left(f_{j t} ,   q_{j t}\right)
$$

\vspace{4mm}

\noindent with $f_{j t} = \mathbf{F}_{j t}^{\prime}\boldsymbol{a_{j t}}$ and $q_{j t} = s_{j, t-1}+\mathbf{F}_{j t}^{\prime} \mathbf{R}_{j t} \mathbf{F}_{j t} $. To make it explicit, we can define as the following manner

\vspace{4mm}

$$
\boldsymbol{a_{j t}} =\left(\begin{array}{c}
a_{j \alpha t} \\
a_{j \beta t}
\end{array}\right) \quad \text { and } \quad \mathbf{R}_{j t} =\left(\begin{array}{cc}
R_{j \alpha t} & R_{j \alpha \beta t} \\
R_{j \alpha \beta t}^{\prime} & R_{j \beta t}
\end{array}\right)
$$

we have

$$
\begin{array}{l}
f_{j t} = x_{j t-1}^{\prime} a_{j \alpha t}+y_{pa(j) t}^{\prime} a_{j \beta t} \\
q_{j t} = s_{j, t-1}+y_{pa(j) t}^{\prime} R_{j \gamma t} y_{pa(j) t} + 2 y_{pa(j) t}^{\prime} R_{j \beta \gamma t}^{\prime} x_{j t-1} + x_{j t-1}^{\prime} R_{j \beta t} x_{j t-1}
\end{array}
$$

\vspace{4mm}

\noindent \textbf{Updating at time t}: with the previous prior, the Normal-Gamma posterior is 

\begin{equation}
\theta_{j t}, \sigma_{j t}^{-1} \mid \mathcal{D}_{t} \sim\mathcal{NG}\left(\boldsymbol{m_{j t}}, \boldsymbol{C_{j t}}, n_{j t}, n_{j t} s_{j, t}\right)
\end{equation}

\vspace{4mm}

\noindent with parameters following standard updating equations:

\vspace{4mm}

Posterior mean vector: \quad \quad \quad \quad \quad \quad \quad \quad \quad $\boldsymbol{m_{j t}}=\boldsymbol{a_{j t}}+\mathbf{A}_{j t} e_{j t} $

Posterior covariance matrix factor: \quad  \quad \quad \quad $\boldsymbol{C_{j t}}=\left(\boldsymbol{R_{j t}}-\mathbf{A}_{j t} \mathbf{A}_{j t}^{\prime} q_{j t}\right) z_{j t} $

Posterior degrees of freedom: \quad \quad \quad \quad \quad \quad $n_{j t}=r_{j t}+1 $

Posterior residual variance estimate:  \quad \quad \quad $s_{j t}=s_{j, t-1} z_{j t} $

\vspace{4mm}
where 
\vspace{4mm}

1 - step ahead forecast error:  \quad  \quad \quad \quad \quad \quad$
e_{j t}=y_{j t}-\boldsymbol{F}_{j t}^{\prime} \boldsymbol{a_{j t}}
$

1-step ahead forecast variance factor: \quad \quad $q_{j t}=s_{j, t-1}+\boldsymbol{F_{j t}^{\prime}} \boldsymbol{R_{j t} F_{j t}}$ 

Adaptive coefficient vector: \quad \quad \quad \quad \quad \quad $\mathbf{A}_{j t}=\boldsymbol{R}_{j t} \boldsymbol{F}_{j t} / q_{j t}
$

Volatility update factor: \quad \quad \quad \quad \quad \quad \quad \quad $ z_{j t}=\left(r_{j t}+e_{j t}^{2} / q_{j t}\right) /\left(r_{j t}+1\right)
$

\clearpage

\section{Appendix B: Joint Predictive Moments}
\label{appendix B}

After computing the predictive density for each equation $j$, we are able to compute the joint predictive density for $\boldsymbol{y_{t}} $ conditional on its risk factor $\textit{parents}$:  ,

\begin{equation}
p\left(\boldsymbol{y}_{t} \mid \mathcal{D}_{t-1}\right)=\prod_{j=1:(K+N)} p\left(y_{j t} \mid \boldsymbol{y}_{pa(j) t}, \mathcal{D}_{t-1}\right)
\end{equation}

\vspace{3mm}

\noindent being simply the product of the already computed $K+N$ different univariate Student's-$t$ distributions. Hence, after series being decoupled for sequential analysis, they are recoupled for multivariate forecasting. The recouple part can be divided in two: one related to the dynamics of the K risk factors and the other related to the N asset returns to be allocated by the investor. We use the predictions from the first part to produce predictions of the second. Therefore, in our decision analysis at Section \ref{sec4: Empirical}, the investor is concerned about the mean and variance of each of this parts for the portfolio allocation study:

\vspace{3mm}

\begin{equation}
\boldsymbol{\lambda}_{t|t-1}=E\left(\mathcal{F}_{t} \mid \mathcal{D}_{t-1}\right) \ \ \ \mbox{and} \ \ \  \boldsymbol{\Sigma}_{t|t-1}^{f}=Var\left(\mathcal{F}_{t} \mid \mathcal{D}_{t-1}\right).
\end{equation}

\noindent for the $K$-vector of expected factor means and the $K \times  K$ expected factor covariance matrix and

\begin{equation}
\boldsymbol{f}_{t|t-1}=E\left(\boldsymbol{r}_{t} \mid \mathcal{D}_{t-1}\right) \ \ \ \mbox{and} \ \ \  \boldsymbol{\Sigma}_{t|t-1}^{r}=Var\left(\boldsymbol{r}_{t} \mid \mathcal{D}_{t-1}\right).
\end{equation}

\noindent for the $N$-vector of expected asset returns and the  $N \times  N$ expected covariance matrix of returns. The structure in Equation (\ref{eq:triangular2}) allows for a recursive computation of moments according to the factor dependency. Since the first dependent variable has an empty \textit{parental set}, the forecast mean and variance for $j = 1$ are given by 

\vspace{3mm}

$$
\begin{aligned}
\lambda_{1t|t-1} &=x_{1 t-1}^{\prime} a_{1 \alpha t} \\
q_{1t|t-1}^{f} &=\frac{r_{1 t}}{r_{1 t}-2}\left(x_{1, t-1}^{\prime} R_{1 \alpha t} x_{1, t-1}+s_{1, t-1}\right)
\end{aligned}
$$

\vspace{3mm}

\noindent inserting $\lambda_{1t|t-1}$ as the first element of $\boldsymbol{\lambda_{t|t-1}}$ and $q_{1t|t-1}^{f} $ the $(1,1)$ element of $\boldsymbol{\Sigma}_{t|t-1}^{f}$. For $j= 2, \ldots, K$, we can recursively find the subsequent predicted moments. Their conditional distributions also follow Student's $t$ distribution, with predictive moments given by

\vspace{3mm}

$$
\begin{aligned}
\lambda_{jt|t-1} &=x_{j t-1}^{\prime} a_{j \alpha t}+\lambda_{pa(j) t|t-1}^{\prime} a_{j \beta t} \\
q_{jt|t-1}^{f} &=\frac{r_{j t}}{r_{j t}-2}\left(s_{j, t-1}+u_{j t}\right) +a_{j \beta t}^{\prime}  \boldsymbol{\Sigma}_{pa(j) t|t-1}^{f} a_{j \beta t}
\end{aligned}
$$

\noindent with $u_{j t}=\lambda_{pa(j) t|t-1}^{\prime} R_{j \beta} \lambda_{pa(j) t|t-1}+\operatorname{tr}\left(R_{j \beta t} \boldsymbol{\Sigma}_{pa(j) t|t-1}^{f}\right)+2 x_{j t-1}^{\prime} R_{j \alpha \beta} \lambda_{pa(j) t|t-1}+x_{j t-1}^{\prime} R_{j \alpha t} x_{j t-1}$. Now, we just need to plug $\lambda_{jt|t-1}$ as the $j$-th element of $\boldsymbol{\lambda_{t|t-1}}$ and $q_{jt}^{f} $ the $(j,j)$ element of $\boldsymbol{\Sigma}_{t|t-1}^{f}$. Finally, the covariance vector among factor $y_{j t}$ and its $\textit{parents}$ $y_{pa(j) t}$ is computed as $C\left(y_{j t}, y_{pa(j), t} \mid \mathcal{D}_{t-1}\right)=\boldsymbol{\Sigma}_{pa(j) t|t-1}^{f} a_{j \beta t} $. Hence, after reaching $j = K$, we have filled all elements of the $K$-vector $\boldsymbol{\lambda_{t|t-1}}$ and the $K \times K$ covariance matrix $\boldsymbol{\Sigma}_{t|t-1}^{f}$.   

For $j= K+1, \ldots, K + N$, we can recursively find the subsequent predicted moments given by:

$$
\begin{aligned}
f_{jt|t-1} &= a_{j \alpha t}+\lambda_{pa(j) t|t-1}^{\prime} a_{j \beta t} \\
q_{jt|t-1}^{r} &=\frac{r_{j t}}{r_{j t}-2}\left(s_{j, t-1}+u_{j t}\right) +a_{j \beta t}^{\prime}  \boldsymbol{\Sigma}_{pa(j) t|t-1}^{f} a_{j \beta t}
\end{aligned}
$$

\noindent with $u_{j t}=\lambda_{pa(j) t|t-1}^{\prime} R_{j \beta} \lambda_{pa(j) t|t-1}+\operatorname{tr}\left(R_{j \beta t} \boldsymbol{\Sigma}_{pa(j) t|t-1}^{f}\right)+2 x_{j t-1}^{\prime} R_{j \alpha \beta} \lambda_{pa(j) t|t-1}+x_{j t-1}^{\prime} R_{j \alpha t} x_{j t-1}$. The $K$-vector of predicted risk factors means, $\lambda_{pa(j) t|t-1}$, the $K$-vector of predicted mean factor loadings, $ a_{j \beta t}$, and the $K\times K$ predicted factor covariance matrix, $\boldsymbol{\Sigma}_{pa(j) t|t-1}^{f}$, may all be filled with zeroes when those elements were not selected from the DMS procedure.

We then plug $f_{jt|t-1}$ as the $(j-K)$-th element of $\boldsymbol{f_{t|t-1}}$ and $q_{jt}^{r} $ the $(j-K, j-K)$ element of $\boldsymbol{\Sigma}_{t|t-1}^{r}$. After reaching $j = K + N$, we have filled all elements of the $N$-vector $\boldsymbol{f_{t|t-1}}$ and all diagonal elements of $\boldsymbol{\Sigma}_{t|t-1}^{r}$. Now we just need to fill all the off-diagonal elements of $\boldsymbol{\Sigma}_{t|t-1}^{r}$ by the off-diagonal elements of the following matrix:

$$
\boldsymbol{\beta_{t|t-1}} \boldsymbol{\Sigma}_{t|t-1}^{f} \boldsymbol{\beta_{t|t-1}}^{\prime}
$$

\noindent where $\boldsymbol{\beta_{t|t-1}}$ is a $N \times K$ matrix containing all predicted mean factor loading that also may be filled with zeroes for those factors not selected by the DMS procedure. 

Finally, we denote $\boldsymbol{\alpha_{t|t-1}} = (a_{K+1, \alpha, t},\dots, a_{K+N, \alpha, t})$ as the $N$-vector containing the time-varying intercepts of each asset return. Then, it can be noticed that the vector of predicted mean asset returns can be represented by:

$$
\boldsymbol{f}_{t|t-1} = \boldsymbol{\alpha_{t|t-1}} + \boldsymbol{\beta}_{t|t-1} \boldsymbol{\lambda}_{t|t-1}
$$

\clearpage

\section{Appendix C: Additional Results}
\label{appendix C}

\begin{figure}[!h]
  \centering
  
  \begin{minipage}[b]{0.47\textwidth}
    \includegraphics[width=\textwidth]{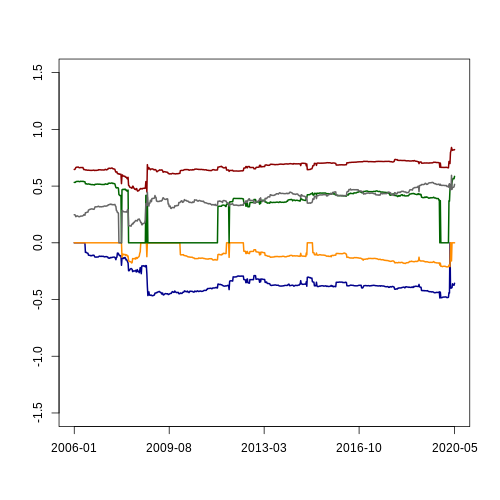}
  \end{minipage}
  \begin{minipage}[b]{0.47\textwidth}
    \includegraphics[width=\textwidth]{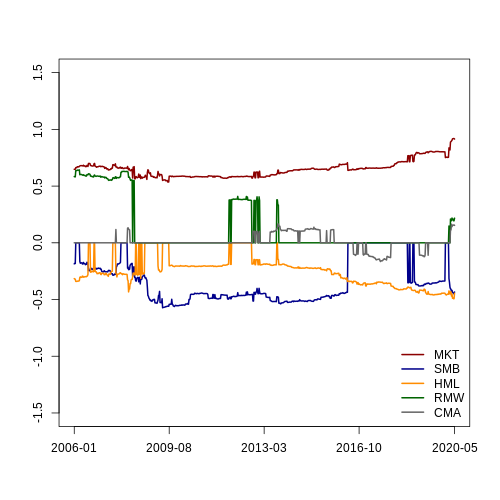}
  \end{minipage}
   \caption{\textit{Time-varying posterior mean factor loadings. Left panel: Coca-Cola. Right Panel: Abbott.}}
   \label{tv_loadings}
\end{figure}

\begin{figure}[!h]
\begin{center}
	\includegraphics[width=15.7cm]{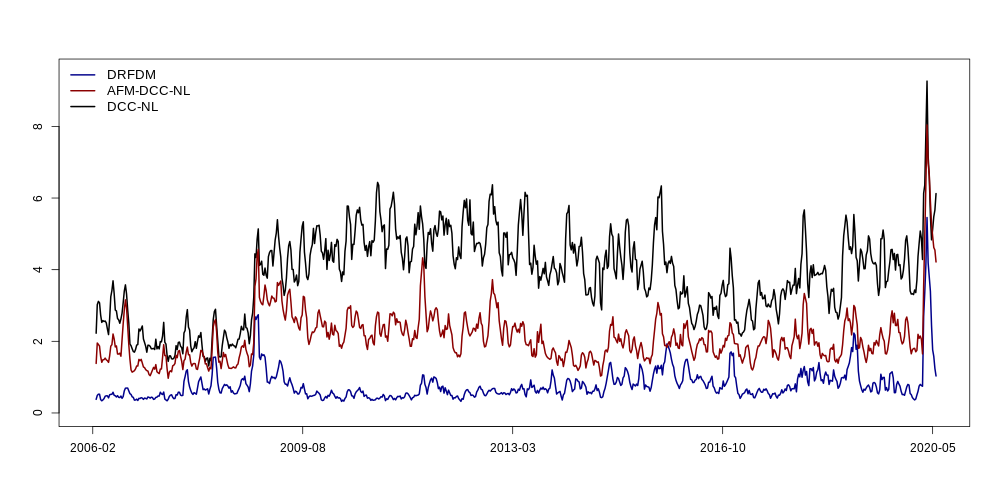} 
\caption{\textit{Out-of-sample turnovers of mean-variance portfolios. The Figure shows rolling sums of weekly turnovers. We use a window size of four weeks as a proxy for monthly turnovers. }}
\label{turnovers}
\end{center}
\end{figure}

\break

\begin{table}[!htbp] \centering 
  \caption{Mean-Variance Portfolio performances (TC = 5 bps and $\tau=15\%)$} 
  \small
  \label{eff_15_tau} 
\begin{tabular}{@{\extracolsep{5pt}}l cccccc} 
\\[-1.8ex]\hline 
\hline \\[-1.8ex] 
 & Turnover & Mean & SD & SR & $\Phi$ \\ 
\hline \\[-1.8ex] 
EFM & $0.220$ & $7.2$ & $13.4$ & $0.540$ & $183.7$ \\ 
AFM-DCC-NL & $0.530$ & $6.7$ & $11.4$ & $0.590$ & $382$ \\ 

DCC-NL & $0.970$ & $6.2$ & $12.3$ & $0.510$ & $217.6$ \\ 
LW & $0.230$ & $7.1$ & $12$ & $0.6$ & $356.6$ \\ 
EWMA(0.99) & $0.830$ & $4.5$ & $14.2$ & $0.320$ & $-201.4$ \\ 
EWMA(0.97) & $2.140$ & $6.9$ & $20$ & $0.350$ & $-928.4$ \\ 
DRFDM & $0.2$ & $11.1$ & $13.0$ & $0.85$ & $636.6$ \\ 

DRFDM (Mom signal) & $0.24$ & $10.4$ & $12.6$ & $0.820$ & $614.5$ \\ 
DRFDM (No Spars.) & $0.13$ & $11.2$ & $14.0$ & $0.80$ & $497.1$ \\ 
TVB-FSV-CV & $0.160$ & $9.6$ & $13.5$ & $0.710$ & $405.2$ \\ 

TVB-FCV-CV & $0.160$ & $9.7$ & $13.6$ & $0.710$ & $407.3$ \\ 
TVB-FSV-SV & $0.170$ & $10.2$ & $13.2$ & $0.770$ & $515.2$ \\ 
TVB-FCV-SV & $0.170$ & $10.2$ & $13.2$ & $0.770$ & $518.8$ \\ 
CB-FSV-SV & $0.160$ & $9.8$ & $13.1$ & $0.750$ & $485$ \\ 
CB-FSV-CV & $0.150$ & $9.1$ & $13.5$ & $0.680$ & $360.5$ \\ 
CB-FCV-CV & $0.150$ & $9.2$ & $13.5$ & $0.680$ & $362$ \\ 
CB-FCV-SV & $0.170$ & $9.8$ & $13.1$ & $0.750$ & $488.3$ \\ 
Factor-W-DLM & $0.7$ & $5.4$ & $13.8$ & $0.390$ & $-54$ \\ 
W-DLM & $0.670$ & $5.7$ & $13.6$ & $0.420$ & $0$ \\ 
EW & $0.030$ & $7.1$ & $21.8$ & $0.320$ & $-1,161$ \\ 
S\&P & $$ & $8.3$ & $18.2$ & $0.460$ & $-486.6$ \\ 
\hline \\[-1.8ex] 
\end{tabular} 

\begin{tablenotes}
      \small 
    \item    \textit{The Table reports out-of-sample portfolio performances for several covariance models following optimal portfolio weights from Equation (\ref{w_eff}) and using an annualized return target of $\tau=15\%$. Mean excess returns (Mean), volatilities (SD), Sharpe Ratios (SR) and management fees ($\Phi$) are reported in annual terms whereas turnover are in weekly terms. Results are computed using returns net of transaction costs of 5 bps. Annualized management fees are computed considering a relative risk aversion of $\gamma=10$. 
}
    \end{tablenotes}
\end{table}

\begin{table}[!htbp] \centering 
  \caption{Mean-Variance Portfolio performances (TC = 5 bps and $\tau=20\%$)} 
  \small
  \label{eff_20_tau} 
\begin{tabular}{@{\extracolsep{5pt}}l cccccc} 
\\[-1.8ex]\hline 
\hline \\[-1.8ex] 
 & Turnover & Mean & SD & SR & $\Phi$ \\ 
\hline \\[-1.8ex] 
EFM & $0.230$ & $7.6$ & $13.5$ & $0.560$ & $274.9$ \\ 

AFM-DCC-NL & $0.540$ & $6.8$ & $11.5$ & $0.590$ & $455.6$ \\ 

DCC-NL & $0.980$ & $6.3$ & $12.4$ & $0.510$ & $291.8$ \\ 
LW & $0.250$ & $7.2$ & $12.1$ & $0.590$ & $424.5$ \\ 
EWMA(0.99) & $0.870$ & $4.3$ & $14.2$ & $0.3$ & $-156.5$ \\ 
EWMA(0.97) & $2.220$ & $6.7$ & $20$ & $0.330$ & $-891.5$ \\ 
DRFDM & $0.220$ & $11.6$ & $13.6$ & $0.860$ & $687.9$ \\

DRFDM (Mom signal) & $0.250$ & $10.7$ & $12.8$ & $0.840$ & $701.1$ \\ 
DRFDM (No Spars.)& $0.15$ & $11.7$ & $14.8$ & $0.79$ & $520.8$ \\ 
TVB-FSV-CV & $0.180$ & $10.4$ & $14.2$ & $0.730$ & $467.2$ \\ 
TVB-FCV-CV & $0.180$ & $10.4$ & $14.2$ & $0.730$ & $469.7$ \\ 
TVB-FSV-SV & $0.190$ & $10.8$ & $13.7$ & $0.780$ & $576.7$ \\ 
TVB-FCV-SV & $0.190$ & $10.8$ & $13.7$ & $0.790$ & $581.2$ \\ 
CB-FSV-SV & $0.170$ & $9.9$ & $13.4$ & $0.740$ & $529.5$ \\ 
CB-FSV-CV & $0.170$ & $9.5$ & $14$ & $0.680$ & $413.4$ \\ 
CB-FCV-CV & $0.170$ & $9.6$ & $14$ & $0.680$ & $415.4$ \\ 
CB-FCV-SV & $0.180$ & $9.9$ & $13.4$ & $0.740$ & $533.5$ \\ 
Factor-W-DLM & $0.730$ & $5.4$ & $14.3$ & $0.380$ & $-51.7$ \\ 
W-DLM & $0.7$ & $5.7$ & $14.1$ & $0.4$ & $0$ \\ 
EW & $0.030$ & $7.1$ & $21.8$ & $0.320$ & $-1,093.1$ \\ 
S\&P & $$ & $8.3$ & $18.2$ & $0.460$ & $-414.7$ \\ 
\hline \\[-1.8ex] 
\end{tabular} 
\begin{tablenotes}
      \small 
    \item    \textit{The Table reports out-of-sample portfolio performances for several covariance models following optimal portfolio weights from Equation (\ref{w_eff}) and using an annualized return target of $\tau=20\%$. Mean excess returns (Mean), volatilities (SD), Sharpe Ratios (SR) and management fees ($\Phi$) are reported in annual terms whereas turnover are in weekly terms. Results are computed using returns net of transaction costs of 5 bps. Annualized management fees are computed considering a relative risk aversion of $\gamma=10$. 
}
    \end{tablenotes}
\end{table}

\begin{table}[!htbp] \centering 
  \caption{Annualized Management Fee ($\Phi$)} 
  \label{fees} 
\begin{tabular}{@{\extracolsep{5pt}}l ccc} 
\\[-1.8ex]\hline 
\hline \\[-1.8ex] 
 &$\Phi (\gamma=2$) & $\Phi (\gamma=6$) \\ 
\hline \\[-1.8ex] 
EFM & $121.4$ & $127.7$ \\ 
AFM-DCC-NL & $139.5$ & $245.7$ \\ 
DCC-NL & $71.2$ & $129.1$ \\ 
LW & $173.8$ & $252.2$ \\ 
EWMA(0.99) & $-123.6$ & $-164$ \\ 
EWMA(0.97) & $-74.1$ & $-505.6$ \\ 
DRFDM & $509.8$ &$546.2$ \\ 
DRFDM (No Spars.) & $498.7$ & $482.2$ \\ 
TVB-FSV-CV & $321.3$ & $324.5$ \\ 
TVB-FCV-CV & $323.3$ & $326.4$ \\ 
TVB-FSV-SV & $403.9$ & $427.6$ \\ 
TVB-FCV-SV & $406.4$ & $430.2$ \\ 
CB-FSV-SV & $406.6$ & $430.3$ \\ 
CB-FSV-CV & $310.9$ & $310.2$ \\ 
CB-FCV-CV & $312.4$ & $311.6$ \\ 
CB-FCV-SV & $408.8$ & $432.8$ \\ 
Factor-W-DLM & $-35.3$ & $-46.7$ \\ 
W-DLM & $0$ & $0$ \\ 
EW & $-35.1$ & $-615.4$ \\ 
S\&P & $107.2$ & $-201$ \\ 
\hline \\[-1.8ex] 
\end{tabular} 

\begin{tablenotes}
      \small 
    \item    \textit{The Table reports out-of-sample annualized management fees ($\Phi$) an investor would be willing to pay to switch from the Wishart Dynamic Linear Model (W-DLM) to the following covariance models. Results refer to optimal porfolios weights from Mean-Variance strategies as in Equation (\ref{w_eff}). Management fees are computed using returns net of transaction costs of 5 bps and relative risk aversions of $\gamma \in \{2, 6 \}$. }
    \end{tablenotes}
    
\end{table}

\begin{table}[!htbp] \centering 
  \caption{Global Minimum Variance Portfolio Performances (TC = 5 bps)} 
  \small
  \label{gmv_tc5} 
  \small
\begin{tabular}{@{\extracolsep{5pt}}l cccccc} 
\\[-1.8ex]\hline 
\hline \\[-1.8ex] 
 & Turnover & Mean & SD & SR & $\Phi$ \\ 
\hline \\[-1.8ex] 
EFM & $0.180$ & $7.4$ & $13.6$ & $0.540$ & $125.8$ \\ 
AFM-DCC-NL & $0.520$ & $7.2$ & $11.4$ & $0.630$ & $391.1$ \\ 

DCC-NL & $0.970$ & $6.9$ & $12.7$ & $0.540$ & $194.9$ \\ 
LW & $0.2$ & $7.2$ & $12$ & $0.6$ & $318.7$ \\ 
EWMA(0.99) & $0.780$ & $5$ & $14.3$ & $0.350$ & $-206.2$ \\ 
EWMA(0.97) & $2.030$ & $7.4$ & $20.2$ & $0.370$ & $-962.2$ \\ 
DRFDM & $0.190$ & $10.7$ & $12.9$ & $0.830$ & $567$ \\ 
DRFDM (No Spars.) & $0.130$ & $11.1$ & $13.9$ & $0.8$ & $462.6$ \\ 
DRFDM ($\alpha=1$)& $0.130$ & $11.3$ & $13.8$ & $0.820$ & $505$ \\  
DRFDM ($\alpha=0.98$)& $0.230$ & $9.2$ & $12.5$ & $0.740$ & $466.2$ \\ 
DRFDM (No Spars.) & $0.13$ & $11.1$ & $13.9$ & $0.80$ & $462.5$ \\ 
3F-DRFDM & $0.140$ & $10.5$ & $13.2$ & $0.790$ & $493.8$ \\ 
4F-DRFDM & $0.170$ & $10.6$ & $13.3$ & $0.8$ & $506.1$ \\ 
6F-DRFDM & $0.210$ & $10.6$ & $12.9$ & $0.820$ & $550.2$ \\ 
TVB-FSV-CV & $0.150$ & $8.6$ & $13.5$ & $0.640$ & $266.7$ \\ 
TVB-FCV-CV & $0.150$ & $8.6$ & $13.5$ & $0.640$ & $267.8$ \\ 
TVB-FSV-SV & $0.170$ & $9.8$ & $13.1$ & $0.750$ & $445.9$ \\ 
TVB-FCV-SV & $0.170$ & $9.8$ & $13.1$ & $0.750$ & $448.1$ \\ 
CB-FSV-SV & $0.160$ & $9.7$ & $13$ & $0.750$ & $445.6$ \\ 
CB-FSV-CV & $0.140$ & $8.7$ & $13.5$ & $0.640$ & $271.1$ \\ 
CB-FCV-CV & $0.140$ & $8.7$ & $13.5$ & $0.640$ & $272.3$ \\ 
CB-FCV-SV & $0.160$ & $9.7$ & $13$ & $0.750$ & $448$ \\ 
Factor-W-DLM & $0.690$ & $5.8$ & $13.7$ & $0.420$ & $-55.3$ \\ 
W-DLM & $0.660$ & $6.1$ & $13.5$ & $0.450$ & $0$ \\ 
EW & $0.030$ & $7.1$ & $21.8$ & $0.320$ & $-1,198.9$ \\ 
S\&P & $$ & $8.3$ & $18.2$ & $0.460$ & $-526.7$ \\ 
\hline \\[-1.8ex] 
\end{tabular} 
\begin{tablenotes}
      \small 
    \item    \textit{The Table reports out-of-sample portfolio performances for several covariance models following optimal Minimum Variance portfolio weights from Equation (\ref{w_gmvp}). Mean excess returns (Mean), volatilities (SD), Sharpe Ratios (SR) and management fees ($\Phi$) are reported in annual terms whereas turnover are in weekly terms. Results are computed using returns net of transaction costs of 5 bps. Annualized management fees are computed considering a relative risk aversion of $\gamma=10$. 
}
    \end{tablenotes}
\end{table}

\begin{table}[!htbp] \centering 
  \caption{Global Minimum Variance Portfolio Performances (TC = 10 bps)} 
  \small
  \label{gmv_tc10} 
\begin{tabular}{@{\extracolsep{5pt}}l cccccc} 
\\[-1.8ex]\hline 
\hline \\[-1.8ex] 
 & Turnover & Mean & SD & SR & $\Phi$ \\ 
\hline \\[-1.8ex] 
EFM & $0.180$ & $6.9$ & $13.6$ & $0.510$ & $250.8$ \\ 

AFM-DCC-NL & $0.520$ & $5.8$ & $11.4$ & $0.510$ & $427.1$ \\ 

DCC-NL & $0.970$ & $4.3$ & $12.7$ & $0.340$ & $111.5$ \\ 
LW & $0.2$ & $6.7$ & $12$ & $0.560$ & $439.7$ \\ 
EWMA(0.99) & $0.780$ & $3$ & $14.2$ & $0.210$ & $-237.4$ \\ 
EWMA(0.97) & $2.030$ & $2.2$ & $20.2$ & $0.110$ & $-1,278.2$ \\ 
DRFDM & $0.190$ & $10.2$ & $12.9$ & $0.790$ & $696.7$ \\ DRFDM (No Spars.)& $0.130$ & $10.8$ & $13.9$ & $0.780$ & $607.9$ \\ 
DRFDM ($\alpha=1$)& $0.130$ & $11$ & $13.8$ & $0.8$ & $649.5$ \\ 
DRFDM ($\alpha=0.98$)& $0.230$ & $8.6$ & $12.4$ & $0.690$ & $582.9$ \\ 
3F-DRFDM & $0.140$ & $10.1$ & $13.2$ & $0.760$ & $634.7$ \\ 
4F-DRFDM & $0.170$ & $10.2$ & $13.3$ & $0.770$ & $639.8$ \\ 
6F-DRFDM & $0.210$ & $10.1$ & $12.9$ & $0.780$ & $674$ \\ 
TVB-FSV-CV & $0.150$ & $8.2$ & $13.5$ & $0.610$ & $403.1$ \\ 
TVB-FCV-CV & $0.150$ & $8.2$ & $13.5$ & $0.610$ & $403.9$ \\ 
TVB-FSV-SV & $0.170$ & $9.4$ & $13.1$ & $0.720$ & $579.4$ \\ 
TVB-FCV-SV & $0.170$ & $9.4$ & $13.1$ & $0.720$ & $581.3$ \\ 
CB-FSV-SV & $0.160$ & $9.3$ & $13$ & $0.710$ & $580.9$ \\ 
CB-FSV-CV & $0.140$ & $8.3$ & $13.5$ & $0.610$ & $408.6$ \\ 
CB-FCV-CV & $0.140$ & $8.3$ & $13.5$ & $0.620$ & $409.5$ \\ 
CB-FCV-SV & $0.160$ & $9.3$ & $13$ & $0.720$ & $583.1$ \\ 
Factor-W-DLM & $0.690$ & $4$ & $13.7$ & $0.290$ & $-63.2$ \\ 
W-DLM & $0.660$ & $4.3$ & $13.5$ & $0.320$ & $0$ \\ 
EW & $0.030$ & $7$ & $21.8$ & $0.320$ & $-1,048.4$ \\ 
S\&P & $$ & $8.3$ & $18.2$ & $0.460$ & $-358$ \\ 
\hline \\[-1.8ex] 
\end{tabular} 
\begin{tablenotes}
      \small 
    \item    \textit{The Table reports out-of-sample portfolio performances for several covariance models following optimal Minimum Variance portfolio weights from Equation (\ref{w_gmvp}). Mean excess returns (Mean), volatilities (SD), Sharpe Ratios (SR) and management fees ($\Phi$) are reported in annual terms whereas turnover are in weekly terms. Results are computed using returns net of transaction costs of 10 bps. Annualized management fees are computed considering a relative risk aversion of $\gamma=10$. 
}
    \end{tablenotes}
\end{table}

\begin{table}[!htbp] \centering 
  \caption{Global Minimum Variance Portfolio Performances (No Transaction Costs)} 
  \small
  \label{gmv_tc0} 
\begin{tabular}{@{\extracolsep{5pt}}l cccccc} 
\\[-1.8ex]\hline 
\hline \\[-1.8ex] 
 & Turnover & Mean & SD & SR & $\Phi$ \\ 
\hline \\[-1.8ex] 
EFM & $0.180$ & $7.9$ & $13.6$ & $0.580$ & $2.7$ \\ 
AFM-DCC-NL & $0.520$ & $8.6$ & $11.4$ & $0.750$ & $355.7$ \\ 
DCC-NL & $0.970$ & $9.4$ & $12.7$ & $0.740$ & $279.4$ \\ 
LW & $0.2$ & $7.8$ & $12$ & $0.650$ & $199.4$ \\ 
EWMA(0.99) & $0.780$ & $7$ & $14.3$ & $0.490$ & $-178.1$ \\ 
EWMA(0.97) & $2.030$ & $12.7$ & $20.2$ & $0.630$ & $-640.4$ \\ 
DRFDM & $0.190$ & $11.2$ & $12.9$ & $0.870$ & $439.3$ \\ 
DRFDM (No Spars.)& $0.130$ & $11.5$ & $13.9$ & $0.820$ & $319.6$ \\ 
DRFDM ($\alpha=1$)& $0.130$ & $11.7$ & $13.8$ & $0.850$ & $362.8$ \\ 
DRFDM ($\alpha=0.98$)& $0.230$ & $9.8$ & $12.5$ & $0.790$ & $351.2$ \\ 
3F-DRFDM & $0.140$ & $10.9$ & $13.2$ & $0.820$ & $355.2$ \\ 
4F-DRFDM & $0.170$ & $11.1$ & $13.3$ & $0.830$ & $374.4$ \\ 
6F-DRFDM & $0.210$ & $11.1$ & $12.9$ & $0.860$ & $428.3$ \\ 
TVB-FSV-CV & $0.150$ & $9$ & $13.5$ & $0.670$ & $132.6$ \\ 
TVB-FCV-CV & $0.150$ & $9$ & $13.5$ & $0.670$ & $133.9$ \\ 
TVB-FSV-SV & $0.170$ & $10.2$ & $13.1$ & $0.780$ & $314.4$ \\ 
TVB-FCV-SV & $0.170$ & $10.3$ & $13.1$ & $0.780$ & $317$ \\ 
CB-FSV-SV & $0.160$ & $10.1$ & $13$ & $0.780$ & $312.3$ \\ 
CB-FSV-CV & $0.140$ & $9$ & $13.5$ & $0.670$ & $135.8$ \\ 
CB-FCV-CV & $0.140$ & $9.1$ & $13.5$ & $0.670$ & $137.3$ \\ 
CB-FCV-SV & $0.160$ & $10.1$ & $13$ & $0.780$ & $315.1$ \\ 
Factor-W-DLM & $0.690$ & $7.6$ & $13.7$ & $0.550$ & $-47.3$ \\ 
W-DLM & $0.660$ & $7.8$ & $13.6$ & $0.570$ & $0$ \\ 
EW & $0.030$ & $7.2$ & $21.8$ & $0.330$ & $-1,346.5$ \\ 
S\&P &  & $8.3$ & $18.2$ & $0.460$ & $-689.1$ \\ 
\hline \\[-1.8ex] 
\end{tabular} 

\begin{tablenotes}
      \small 
    \item    \textit{The Table reports out-of-sample portfolio performances for several covariance models following optimal Minimum Variance portfolio weights from Equation (\ref{w_gmvp}). Mean excess returns (Mean), volatilities (SD), Sharpe Ratios (SR) and management fees ($\Phi$) are reported in annual terms whereas turnover are in weekly terms. Results are reported considering no transaction costs (TC= 0 bp). Annualized management fees are computed considering a relative risk aversion of $\gamma=10$. 
}
    \end{tablenotes}
\end{table}

\break

\begin{table}[!htbp] \centering 
  \caption{Mean-Variance Portfolio Performances ($|w_{i}|\leq 5\%$)} 
  \label{eff_w5} 
\begin{tabular}{@{\extracolsep{5pt}}l cccccc} 
\\[-1.8ex]\hline 
\hline \\[-1.8ex] 
 & Turnover & Mean & SD & SR & $\Phi$ \\ 
\hline \\[-1.8ex] 
EFM & $0.230$ & $6.9$ & $13.3$ & $0.520$ & $154.5$ \\ 
AFM-DCC-NL & $0.510$ & $6.9$ & $11.3$ & $0.610$ & $411.2$ \\ 

DCC-NL & $0.950$ & $6.6$ & $12.3$ & $0.540$ & $259.5$ \\ 
LW & $0.230$ & $7.2$ & $11.9$ & $0.610$ & $371.7$ \\ 
EWMA(0.99) & $0.7$ & $3.9$ & $13$ & $0.3$ & $-105.6$ \\ 
EWMA(0.97) & $1.190$ & $6.7$ & $14$ & $0.480$ & $47.6$ \\ 
DRFDM & $0.190$ & $10.5$ & $12.8$ & $0.820$ & $605.2$ \\ 
DRFDM (No Spars.)& $0.14$ & $10.7$ & $13.8$ & $0.77$ & $479.5$ \\ 
TVB-FSV-CV & $0.150$ & $8.9$ & $13.4$ & $0.660$ & $348.6$ \\ 
TVB-FCV-CV & $0.160$ & $8.9$ & $13.4$ & $0.660$ & $350.4$ \\ 
TVB-FSV-SV & $0.170$ & $9.6$ & $13$ & $0.740$ & $473.1$ \\ 
TVB-FCV-SV & $0.180$ & $9.6$ & $13$ & $0.740$ & $476.1$ \\ 
CB-FSV-SV & $0.170$ & $9.6$ & $13$ & $0.740$ & $475.7$ \\ 
CB-FSV-CV & $0.150$ & $8.8$ & $13.4$ & $0.650$ & $330.3$ \\ 
CB-FCV-CV & $0.150$ & $8.8$ & $13.4$ & $0.650$ & $331.6$ \\ 
CB-FCV-SV & $0.170$ & $9.6$ & $13$ & $0.740$ & $478.4$ \\ 
Factor-W-DLM & $0.620$ & $4.6$ & $12.8$ & $0.360$ & $-3.3$ \\ 
W-DLM & $0.6$ & $4.6$ & $12.7$ & $0.360$ & $0$ \\ 
EW & $0.030$ & $7.1$ & $21.8$ & $0.320$ & $-1,163.5$ \\ 
S\&P &  & $8.3$ & $18.2$ & $0.460$ & $-489.3$ \\ 
\hline \\[-1.8ex] 
\end{tabular} 
\begin{tablenotes}
      \small 
    \item    \textit{The Table reports out-of-sample portfolio performances for several covariance models in a mean-variance problem as Equation (\ref{mvp_opt}) after including a maximum (absolute) weight constraint of 5\%. We use an annualized return target of $\tau=10\%$. Mean excess returns (Mean), volatilities (SD), Sharpe Ratios (SR) and management fees ($\Phi$) are reported in annual terms whereas turnover are in weekly terms. Results are computed using returns net of transaction costs of 5 bps. Annualized management fees are computed considering a relative risk aversion of $\gamma=10$. 
}
    \end{tablenotes}

\end{table}

\begin{table}[!htbp] \centering 
  \caption{Global Minimum Variance Portfolio Performances ($|w_{i}|\leq 5\%$)} 
  \label{gmv_w5} 
\begin{tabular}{@{\extracolsep{5pt}}l cccccc} 
\\[-1.8ex]\hline 
\hline \\[-1.8ex] 
 & Turnover & Mean & SD & SR & $\Phi$ \\ 
\hline \\[-1.8ex] 
EFM& $0.180$ & $7.4$ & $13.6$ & $0.540$ & $136.2$ \\ 

AFM-DCC-NL & $0.510$ & $7.5$ & $11.4$ & $0.660$ & $439.1$ \\ 

DCC-NL & $0.950$ & $7.3$ & $12.6$ & $0.580$ & $254.1$ \\ 
LW & $0.2$ & $7.4$ & $12$ & $0.620$ & $346.6$ \\ 
EWMA(0.99) & $0.670$ & $4.3$ & $13.1$ & $0.330$ & $-99.3$ \\ 
EWMA(0.97) & $1.130$ & $7.6$ & $14$ & $0.550$ & $107.2$ \\ 
DRFDM & $0.190$ & $10.7$ & $12.9$ & $0.830$ & $579.1$ \\ 
DRFDM (No Spars.)& $0.13$ & $11.1$ & $14.0$ & $0.80$ & $466.9$ \\ 
TVB-FSV-CV & $0.150$ & $8.6$ & $13.5$ & $0.640$ & $277.4$ \\ 
TVB-FCV-CV & $0.150$ & $8.6$ & $13.5$ & $0.640$ & $278.5$ \\ 
TVB-FSV-SV & $0.170$ & $9.8$ & $13.1$ & $0.750$ & $457.5$ \\ 
TVB-FCV-SV & $0.170$ & $9.8$ & $13.1$ & $0.750$ & $459.8$ \\ 
CB-FSV-SV & $0.160$ & $9.7$ & $13$ & $0.750$ & $456.9$ \\ 
CB-FSV-CV & $0.140$ & $8.7$ & $13.5$ & $0.640$ & $281.6$ \\ 
CB-FCV-CV & $0.140$ & $8.7$ & $13.5$ & $0.640$ & $282.8$ \\ 
CB-FCV-SV & $0.160$ & $9.7$ & $13$ & $0.750$ & $459.4$ \\ 
Factor-W-DLM & $0.620$ & $5$ & $12.9$ & $0.390$ & $-8.7$ \\ 
W-DLM & $0.6$ & $5$ & $12.8$ & $0.390$ & $0$ \\ 
EW & $0.030$ & $7.1$ & $21.8$ & $0.320$ & $-1,189.3$ \\ 
S\&P &  & $8.3$ & $18.2$ & $0.460$ & $-516.6$ \\ 
\hline \\[-1.8ex] 
\end{tabular} 

\begin{tablenotes}
      \small 
    \item    \textit{The Table reports out-of-sample portfolio performances for several covariance models in a global minimum-variance optimization problem as in Equation (\ref{min_var}) after including a maximum (absolute) weight constraint of 5\%. Mean excess returns (Mean), volatilities (SD), Sharpe Ratios (SR) and management fees ($\Phi$) are reported in annual terms whereas turnover are in weekly terms. Results are computed using returns net of transaction costs of 5 bps. Annualized management fees are computed considering a relative risk aversion of $\gamma=10$. 
}
    \end{tablenotes}
\end{table}

\clearpage

\bibliographystyle{ecta}
\bibliography{referencias}
\clearpage




\end{document}